
\input harvmac

\Title{\vbox{\baselineskip12pt\hbox{McGill/94-55}
\hbox{CERN-TH/95-57}
\hbox{hep-th/9506065}
}}
{\vbox{\centerline {Black Holes and Solitons in String Theory}}}
\centerline{Ramzi R.~Khuri}
\bigskip\centerline{{\it Physics Department}}
\centerline{\it McGill University}\centerline{\it Montreal, PQ,
H3A 2T8 Canada}
\medskip
\centerline{and}
\medskip
\bigskip\centerline{{\it Theory Division, CERN}}
\centerline{\it CH-1211, Geneva 23, Switzerland}
\vskip .3in
In this review I discuss various aspects of some of
the
recently constructed black hole and soliton solutions in
string theory. I begin with the axionic instanton
and related solutions of bosonic and
heterotic string theory. The latter ten-dimensional solutions
can be compactified to supersymmetric monopole, string
and domain wall solutions which break $1/2$ of the spacetime
supersymmetries of $N=4, D=4$ heterotic string theory,
and which can be generalized to two-parameter charged black
hole solutions.
The low-energy dynamics of these solutions is also discussed,
as
well as their connections with strong/weak coupling duality and
target space duality in string theory. Finally, new solutions
 are
presented which break $3/4, 7/8$ and all of the spacetime
supersymmetries and which also arise in more realistic $N=1$ and $N=2$
compactifications.

\Date{\vbox{\baselineskip12pt\hbox{McGill/94-55}
\hbox{CERN-TH/95-57}
\hbox{June 1995}
}}

\def\sqr#1#2{{\vbox{\hrule height.#2pt\hbox{\vrule width
.#2pt height#1pt \kern#1pt\vrule width.#2pt}\hrule height.#2pt}}}
\def\Box{\mathchoice\sqr64\sqr64\sqr{4.2}3\sqr33}

\def\rijkl{R^i{}_{jkl}}
\def\grijkl{\hat R^i{}_{jkl}}

\def\met {g_{mn}}

\def\b#1{\vec\beta_{#1}}
\def\x#1{\vec X_{#1}}

\def\a{\alpha}
\def\x{{(R^2-2)v^2\over 1-v^2}}
\lref\prep{M. J. Duff, R. R. Khuri and J. X. Lu,
 NI-94-017, CTP/TAMU-67/92,
McGill/94/53, CERN-TH.7542/94, hepth/9412184
(to appear in Physics Reports).}

\lref\fabjr{M. Fabbrichesi, R. Jengo and K. Roland, Nucl.
Phys. {\bf B402} (1993) 360.}

\lref\jxthesis{J. X. Lu, {\it Supersymmetric Extended
Objects}, Ph.D. Thesis, Texas A\&M University (1992),
UMI {\bf 53 08B}, Feb. 1993.}

\lref\coleman{S. Coleman, ``Classical Lumps and Their Quantum
Descendants'', in {\it New Phenomena in Subnuclear Physics},
ed A. Zichichi (Plenum, New York, 1976).}

\lref\jackiw{R. Jackiw, Rev. Mod. Phys. {\bf 49} (1977) 681.}

\lref\hink{M. B. Hindmarsh and T. W. B. Kibble, SUSX-TP-94-74,
IMPERIAL/TP/94-95/5, NI 94025, hepph/9411342.}

\lref\senmod{A. Sen, Mod. Phys. Lett. {\bf A8} (1993) 2023.}

\lref\kikyam{K. Kikkawa and M. Yamasaki, Phys. Lett. {\bf B149}
(1984) 357.}

\lref\sakai{N. Sakai and I. Senda, Prog. Theor. Phys. {\bf 75}
(1986) 692.}

\lref\bush{T. Busher, Phys. Lett. {\bf B159} (1985) 127.}

\lref\nair{V. Nair, A. Shapere, A. Strominger and F. Wilczek,
Nucl. Phys. {\bf B322} (1989) 167.}

\lref\duff{M. J. Duff, Nucl. Phys. {\bf B335} (1990)
610.}

\lref\tseyv{A. A. Tseytlin and C. Vafa, Nucl. Phys. {\bf B372}
(1992) 443,}

\lref\tseycqg{A. A. Tseytlin, Class. Quantum Grav. {\bf 9}
(1992) 979.}

\lref\givpr{A. Giveon, M. Porrati and E. Rabinovici,
(to appear in Phys. Rep. C).}

\lref\gibkal{G. W. Gibbons and R. Kallosh, NI-94003,
hepth/9407118.}

\lref\vafw{C. Vafa and E. Witten, Nucl. Phys. {\bf B431}
(1994) 3.}

\lref\seiw{N. Seiberg and E. Witten, Nucl. Phys. {\bf B426}
(1994) 19.}

\lref\seiwone{N. Seiberg and E. Witten, Nucl. Phys. {\bf B431}
(1994) 484.}

\lref\cerdfv{A. Ceresole, R. D'Auria, S. Ferrara and
A. Van Proeyen, CERN-TH 7510/94, POLFIS-TH.08/94,
UCLA 94/TEP/45, KUL-TF-94/44, hepth/9412200.}

\lref\cerdfvone{A. Ceresole, R. D'Auria, S. Ferrara and
A. Van Proeyen, CERN-TH 7547/94, POLFIS-TH.01/95,
UCLA 94/TEP/45, KUL-TF-95/4, hepth/9502072.}

\lref\gauh{J. Gauntlett and J. H. Harvey, EFI-94-30,
hepth/9407111.}

\lref\frak{P. H. Frampton and T. W. Kephart, IFP-708-UNC,
VAND-TH-94-15.}

\lref\hawhr{S. W. Hawking, G. T. Horowitz and S. F. Ross,
NI-94-012, DAMTP/R 94-26, UCSBTH-94-25, gr-qc/9409013.}

\lref\ellmn{J. Ellis, N. E. Mavromatos and D. V. Nanopoulos,
Phys. Lett. {\bf B278} (1992) 246.}

\lref\kaln{S. Kalara and N. Nanopoulos, Phys. Lett. {\bf B267}
(1992) 343.}

\lref\glasgow{M. J. Duff, NI-94-016, CTP-TAMU-48/94,
hepth/9410210.}

\lref\dufnew{M. J. Duff, NI-94-033, CTP-TAMU-49/94,
hepth/9501030.}

\lref\bhs{R. R. Khuri, Helv. Phys. Acta {\bf 67} (1994) 884.}

\lref\cecfg{S. Cecotti, S. Ferrara and L. Girardello,
Int. J. Mod. Phys. {\bf A4} (1989) 2475.}

\lref\ferquat{S. Ferrara and S. Sabharwal,
Nucl. Phys. {\bf B332} (1990) 317.}

\lref\fere{R. C. Ferrell and D. M. Eardley,
Phys. Rev. Lett. {\bf 59} (1987) 1617.}

\lref\reyt{S. J. Rey and T. R. Taylor, Phys. Rev. Lett. {\bf 71}
(1993) 1132.}

\lref\senzwione{A. Sen and B. Zwiebach, Nucl. Phys.
{\bf B414} (1994) 649.}

\lref\senzwitwo{A. Sen and B. Zwiebach, Nucl. Phys.
{\bf B423} (1994) 580.}

\lref\bddo{T. Banks, A. Dabholkar, M. R. Douglas and
M. O'Loughlin, Phys. Rev. {\bf D45} (1992) 3607.}

\lref\bos{T. Banks, M. O'Loughlin and A. Strominger,
Phys. Rev. {\bf D47} (1993) 4476.}

\lref\kir{E. Kiritsis, Nucl. Phys. {\bf B405} (1993) 109.}

\lref\dufr{M. J. Duff and J. Rahmfeld, Phys. Lett. {\bf B345}
(1995) 441.}

\lref\duffkk{M. J. Duff, NI-94-015, CTP-TAMU-22/94,
hepth/9410046.}

\lref\bakone{I. Bakas, Nucl. Phys. {\bf B428} (1994) 374.}

\lref\baktwo{I. Bakas, Phys. Lett. {\bf B343} (1995) 103.}

\lref\baksfet{I. Bakas and K. Sfetsos, CERN-TH-95-16,
hepth/9502065.}

\lref\back{C. Bachas and E. Kiritsis, Phys. Lett. {\bf B325}
(1994) 103.}

\lref\bko{E. Bergshoeff, R. Kallosh and T. Ortin,
UG-8/94, SU-ITP-94-19, QMW-PH-94-13, hepth/9410230.}

\lref\vilsh{A. Vilenkin and E. P. Shellard,
{\it Cosmic String and Other Topological Defects},
(Cambridge University Press, 1994).}

\lref\bfrm{M. Bianchi, F. Fucito, G. C. Rossi and M. Martellini,
hepth/9409037.}

\lref\hult{C. M. Hull and P. K. Townsend, QMW-94-30,
R/94/33, hepth/9410167.}

\lref\ght{G. W. Gibbons, G. T. Horowitz and P. K. Townsend,
R/94/28, UCSBTH-94-35, hepth/9410073.}

\lref\dufkmr{M. J. Duff, R. R. Khuri, R. Minasian and
J. Rahmfeld, Nucl. Phys. {\bf B418} (1994) 195.}

\lref\sentd{A. Sen, Nucl. Phys. {\bf B434} (1995) 179.}

\lref\maha{J. Maharana, NI-94023, hepth/9412235.}

\lref\gresss{M. B. Green and J. Schwarz, Phys. Lett. {\bf B136}
(1984) 367.}

\lref\sie{W. Siegel, Phys. Lett. {\bf B128} (1983) 397.}

\lref\dir {P. A. M. Dirac, Pro. R. Soc. {\bf A133} (1931) 60.}

\lref\tho {G. t'Hooft, Nucl. Phys. {\bf B79} (1974) 276.}

\lref\pol {A. M. Polyakov, Sov. Phys. JETP Lett. {\bf 20}
(1974) 194.}

\lref\mono {C. Montonen and D. Olive, Phys. Lett. {\bf B72}
(1977) 117.}

\lref\col {S. Coleman, Phys. Rev. {\bf D11} (1975) 2088.}

\lref\grohmr {D. J. Gross, J. A. Harvey, E. Martinec and
 R. Rohm,
Nucl. Phys. {\bf B256} (1985) 253.}

\lref\ginone {P. Ginsparg,
 {\it Conformal Field Theory}, Lectures given
at Trieste Summer School, Trieste, Italy, 1991.}

\lref\calhstwo {C. Callan, J. Harvey and A. Strominger,
 {\it Supersymmetric String Solitons}, Lectures given
at Trieste Summer School, Trieste, Italy, 1991.}

\lref\sch {J. H. Schwarz, {\it Supersymmetry and Its
 Applications}
ed G. W. Gibbons {\it et al} (Cambridge University Press,
1986).}

\lref\huglp {J. Hughes, J. Liu and J. Polchinski, Phys. Lett.
{\bf B180} (1986)
370.}

\lref\berst {E. Bergshoeff, E. Sezgin and P. K. Townsend,
 Phys. Lett. {\bf B189} (1987) 75.}

\lref\achetw {A. Achucarro, J. Evans, P. K.  Townsend and
D. Wiltshire,
 Phys.
Lett. {\bf B198} (1987) 441.}

\lref\belpst {A. A. Belavin, A. M. Polyakov, A. S. Schwartz and
Yu. S. Tyupkin,
Phys. Lett. {\bf B59} (1975) 85.}

\lref\oset {D. O'Se and D. H. Tchrakian, Lett. Math. Phys.
{\bf 13} (1987) 211.}

\lref\groks {B. Grossman, T. W. Kephart and J. D. Stasheff,
 Commun. Math. Phys. {\bf 96} (1984) 431;
Commun. Math. Phys. {\bf 100} (1985) 311.}

\lref\grokstwo {B. Grossman, T. W. Kephart and J. D. Stasheff,
 Phys. Lett. {\bf B220}
 (1989)  431.}

\lref\tch{D. H. Tchrakian, Phys. Lett. {\bf B150}  (1985)  360.}

\lref\fubn{S. Fubini and H. Nicolai,
 Phys. Lett. {\bf B155}  (1985)  369}

\lref\fain{D. B. Fairlie and J. Nuyts,
 J. Phys. {\bf A17}  (1984)  2867.}

\lref\str {A. Strominger,  Nucl. Phys. {\bf B343}
(1990) 167.}

\lref\duflhs {M. J. Duff and J. X. Lu,
Phys. Rev. Lett. {\bf 66} (1991) 1402.}

\lref\godo {P. Goddard and D. Olive, Rep. Prog. Phys. {\bf 41}
(1978) 1357.}

\lref\colone {S. Coleman, Proc. 1975 Int. School on Subnuclear
Physics,
Erice, ed A. Zichichi  (Plenum, New York, 1977); Proc. 1981
Int. School
on Subnuclear Physics, Erice, ed A. Zichichi (Plenum, New York,
1983).}

\lref\wuy {T. T. Wu and C. N. Yang, Nucl. Phys. {\bf B107}
(1976) 365.}

\lref\dufldl {M. J. Duff and J. X. Lu, Class. Quantum Grav.
{\bf 9} (1992) 1.}

\lref\tei {C. Teitelboim,
Phys. Lett. {\bf B167} (1986) 69.}

\lref\nep {R. I. Nepomechie,
Phys. Rev. {\bf D31} (1984) 1921.}

\lref\raj {R. Rajaraman, {\it Solitons and Instantons}
(North--Holland, Amsterdam, 1982).}

\lref\wito {E. Witten and D. Olive, Phys. Lett.
{\bf B78} (1978) 97.}

\lref\pras {M. K. Prasad and C. M. Sommerfield, Phys. Rev. Lett.
{\bf 35} (1975) 760.}

\lref\corg {E. Corrigan and P. Goddard, Commun. Math. Phys.
{\bf 80} (1981)
575.}

\lref\godno {P. Goddard, J. Nuyts and D. Olive, Nucl. Phys.
{\bf B125} (1977)
1.}

\lref\osb {H. Osborn, Phys. Lett. {\bf B83} (1979) 321.}

\lref\egugh {T. Eguchi, P. B. Gilkey and A. J. Hanson,
Phys. Rep. {\bf 66}
(1980) 213.}

\lref\corf {E. F. Corrigan and D. B. Fairlie, Phys. Lett.
{\bf B67} (1977) 69.
}

\lref\atidhm {M. F. Atiyah, V. G. Drinfeld, N. J. Hitchin and
Y. I. Manin,
Phys. Lett. {\bf A65} (1978) 185.}

\lref\dun {A. R. Dundarer,
 Mod. Phys. Lett. {\bf A5} (1991) 409.}

\lref\cha {A. H. Chamseddine, Phys. Rev. {\bf D24} (1981) 3065.}

\lref\berrwv {E. A. Bergshoeff, M. de Roo, B. de Wit and
 P. van Nieuwenhuizen,
Nucl. Phys. {\bf B195}  (1982)  97}

\lref\cham{G. F. Chapline and N. S. Manton, Phys. Lett.
{\bf B120} (1983) 105.}

\lref\gatn {S. J. Gates and H. Nishino, Phys. Lett. {\bf B173}
(1986) 52.}

\lref\sala{A. Salam and E. Sezgin, Physica Scripta {\bf 32}
(1985) 283.}

\lref\duf {M. J. Duff,  Class.
 Quantum  Grav. {\bf 5} (1988) 189.}

\lref\duflfb {M. J. Duff and J. X. Lu, Nucl. Phys. {\bf B354}
(1991) 141.}

\lref\dabghr {A. Dabholkar, G. W. Gibbons, J. A. Harvey and
F. Ruiz Ruiz,
 Nucl. Phys. {\bf B340} (1990) 33.}

\lref\berdps {E. Bergshoeff, M. J. Duff, C. N. Pope and
E. Sezgin,
 Phys. Lett.
{\bf B199} (1987) 69.}

\lref\tow {P. K. Townsend,
Phys. Lett. {\bf B202} (1988) 53.}

\lref\dufhis {M. J. Duff, P. S. Howe, T. Inami and K. Stelle,
 Phys. Lett.
{\bf B191} (1987) 70.}

\lref\dufs {M. J. Duff and K. Stelle, Phys. Lett. {\bf B253}
(1991) 113.}

\lref\berst {E. Bergshoeff, E. Sezgin and P. K. Townsend, Ann.
 Phys. {\bf 199}
(1990) 340.}

\lref\duflrsfd {M. J. Duff and J. X. Lu, Nucl. Phys. {\bf B354}
(1991) 129.}

\lref\gresone {M. Green and J. Schwarz, Phys. Lett. {\bf B151}
(1985) 21.}

\lref\bercgw {C. W. Bernard, N. H. Christ, A. H. Guth and
E. J. Weinberg,
 Phys. Rev.
 {\bf D16} (1977) 2967.}

\lref\hars {J. Harvey and A. Strominger,
 Phys. Rev. Lett. {\bf 66} (1991) 549.}

\lref\gres {M. Green and J. Schwarz, Phys. Lett. {\bf B149}
(1984) 117.}

\lref\elljm {J. Ellis, P. Jetzer and L. Mizrachi,
Nucl. Phys. {\bf B303} (1988)
1.}

\lref\dixds {J. Dixon, M. J. Duff and E. Sezgin, Phys. Lett.
{\bf B279} (1992) 265.}

\lref\berrs {E. Bergsheoff, M. Rakowski and E. Sezgin, Phys.
 Lett. {\bf
B185}  (1987)  371}

\lref\berd{E. Bergsheoff and M. de Roo, Nucl. Phys. {\bf B328}
(1989)
439}

\lref\dersw{M. de Roo, H. Suelmann and A. Wiedemann, preprint
UG--1/92  (1992).}

\lref\gresw {M. Green, J. Schwarz and E. Witten,
{\it Superstring
theory} (Cambridge University Press, 1987).}

\lref\duflloop {M. J. Duff and J. X. Lu, Nucl. Phys. {\bf B357}
  (1991)  534.}

\lref\ven {G. Veneziano,
Europhys. Lett. {\bf 2}  (1986)  199.}

\lref\cain {Y. Cai and C. A. Nunez, Nucl. Phys. {\bf B287}
(1987)  41}

\lref\gros{D. J. Gross and J. Sloan, Nucl. Phys. {\bf B291}
(1987)  41.}

\lref\ellm {J. Ellis and L. Mizrachi,
  Nucl. Phys. {\bf B327}  (1989)  595.}

\lref\calfmp {C. G. Callan, D. Friedan, E. J. Martinec and
M. J. Perry,
Nucl. Phys. {\bf B262}  (1985)  593.}

\lref\grestwo {M. Green and J. Schwarz, Phys. Lett. {\bf B173}
 (1986)  52.}

\lref\lin {U. Lindstrom, in Supermembranes and Physics in 2 + 1
Dimensions, ed. M. J. Duff, C. N. Pope and E. Sezgin
(World Scientific,
Singapore) (1990).}

\lref\dufone {M. J. Duff, Class. Quantum Grav. {\bf 6}  (1989)
1577.}

\lref\callny {C. Callan, C. Lovelace, C. Nappi and S. Yost,
Nucl. Phys.
{\bf B308}  (1988)  221.}

\lref\frat {E. Fradkin and A. Tseytlin, Phys. Lett. {\bf B158}
(1985)  316.}

\lref\duflselft {M. J. Duff and J. X. Lu, Phys. Lett.
{\bf B273}  (1991)  409.}

\lref\hors {G. Horowitz and A. Strominger, Nucl. Phys.
{\bf B360}  (1991) 197. }

\lref\witone {E. Witten,  Phys. Lett. {\bf B86}  (1979)
283.}

\lref\schone {J. Schwarz, Nucl. Phys. {\bf B226}  (1983)  269.}

\lref\zwa {D. Zwanziger, Phys. Rev. {\bf 176}  (1968)  1480,
1489.}

\lref\schwing {J. Schwinger, Phys. Rev. {\bf 144} (1966) 1087;
{\bf 173}
(1968) 1536.}

\lref\gibt{G.W. Gibbons and P.K. Townsend, Phys. Rev. Lett.
{\bf 71} (1993)
3754.}

\lref\duflblacks {M. J. Duff and J. X. Lu,
 Nucl. Phys. {\bf B416} (1994) 301.}

\lref\dufklsin {M. J. Duff, R. R. Khuri and J. X. Lu,
 Nucl. Phys. {\bf B377}
(1992) 281.}

\lref\calk {C.~G.~Callan and R.~ R.~Khuri,
Phys. Lett. {\bf B261} (1991) 363.}

\lref\gib {G. W. Gibbons, Nucl. Phys. {\bf B207} (1982) 337.}

\lref\gibm {G. W. Gibbons and K. Maeda, Nucl. Phys. {\bf B298}
(1988) 741.}

\lref\rey{S. J. Rey, in Proceedings of Tuscaloosa
Workshop on Particle Physics, (Tuscaloosa, Alabama, 1989).}

\lref\reyone {S. J.~Rey, Phys. Rev. {\bf D43} (1991) 526.}

\lref\antben {I.~Antoniadis, C.~Bachas, J.~Ellis and
 D.~V.~Nanopoulos,
Phys. Lett. {\bf B211} (1988) 393.}

\lref\antbenone {I.~Antoniadis, C.~Bachas, J.~Ellis and
D.~V.~Nanopoulos,
Nucl. Phys. {\bf B328} (1989) 117.}

\lref\mett {R.~R.~Metsaev and A.~A.~Tseytlin, Phys. Lett.
{\bf B191} (1987) 354.}

\lref\mettone {R.~R.~Metsaev and A.~A.~Tseytlin,
Nucl. Phys. {\bf B293} (1987) 385.}

\lref\calkp {C.~G.~Callan,
I.~R.~Klebanov and M.~J.~Perry, Nucl. Phys. {\bf B278} (1986)
78.}

\lref\lov {C.~Lovelace, Phys. Lett. {\bf B135} (1984) 75.}

\lref\friv {B.~E.~Fridling and A.~E.~M.~Van de Ven,
Nucl. Phys. {\bf B268} (1986) 719.}

\lref\gepw {D.~Gepner and E.~Witten, Nucl. Phys. {\bf B278}
(1986) 493.}

\lref\din {M.~Dine, Lectures delivered at
TASI 1988, Brown University (1988) 653.}

\lref\berdone {E.~A.~Bergshoeff and M.~de Roo, Phys. Lett.
{\bf B218} (1989)
210.}

\lref\calhs{C.~G.~Callan, J.~A.~Harvey and A.~Strominger,
Nucl. Phys.
{\bf B359} (1991) 611.}

\lref\calhsone{C.~G.~Callan, J.~A.~Harvey and A.~Strominger,
Nucl. Phys.
{\bf B367} (1991) 60.}

\lref\thoone{G.~'t~Hooft, Phys. Rev. Lett. {\bf 37} (1976) 8.}

\lref\wil{F.~Wilczek, in
{\it Quark confinement and field theory},
Eds. D.~Stump and D.~Weingarten, (John Wiley and Sons, New York,
1977).}

\lref\jacnr{R.~Jackiw, C.~Nohl and C.~Rebbi, Phys. Rev.
{\bf D15} (1977)
1642.}

\lref\khuinst{R.~R.~Khuri, Phys. Lett.
{\bf B259} (1991) 261.}

\lref\khumant{R.~R.~Khuri, Nucl. Phys.
 {\bf B376} (1992) 350.}

\lref\khumono{R.~R.~Khuri,
 Phys. Lett. {\bf B294} (1992) 325.}

\lref\khumonscat{R.~R.~Khuri,
Phys. Lett. {\bf B294} (1992) 331.}

\lref\khumonex{R.~R.~Khuri,
Nucl. Phys. {\bf B387} (1992) 315.}

\lref\khumonin{R.~R.~Khuri,
 Phys. Rev. {\bf D46} (1992) 4526.}

\lref\khugeo{R.~R.~Khuri,
Phys. Lett. {\bf 307} (1993) 302.}

\lref\khuscat{R.~R.~Khuri,
 Nucl. Phys. {\bf B403} (1993) 335.}

\lref\khuwind {R.~R.~Khuri, Phys. Rev. {\bf D48} (1993) 2823.}

\lref\gin{P.~Ginsparg, Lectures delivered at
Les Houches summer session, June 28--August 5, 1988.}

\lref\alljj{R. W. Allen, I. Jack and D. R. T. Jones,
 Z. Phys. {\bf C41}
(1988) 323.}

\lref\sev{A. Sevrin, W. Troost and A. van Proeyen,
Phys. Lett. {\bf B208} (1988) 447.}

\lref\schout{K. Schoutens, Nucl. Phys. {\bf B295} [FS21] (1988)
634.}

\lref\harl{J.~A.~Harvey and J.~Liu, Phys. Lett. {\bf B268}
(1991) 40.}

\lref\man{N.~S.~Manton, Nucl. Phys. {\bf B126} (1977) 525.}

\lref\manone{N.~S.~Manton, Phys. Lett. {\bf B110} (1982) 54.}

\lref\mantwo{N.~S.~Manton, Phys. Lett. {\bf B154} (1985) 397.}

\lref\atihone{M.~F.~Atiyah and N.~J.~Hitchin, Phys. Lett.
{\bf A107}
(1985) 21.}

\lref\atihtwo{M.~F.~Atiyah and N.~J.~Hitchin, {\it The Geometry
and
Dynamics of Magnetic Monopoles}, (Princeton University Press,
1988).}

\lref\polc{J.~Polchinski, Phys. Lett. {\bf B209} (1988) 252.}

\lref\gibhp{G.~W.~Gibbons and S.~W.~Hawking, Phys. Rev.
{\bf D15}
(1977) 2752.}

\lref\gibhpone{G.~W.~Gibbons, S.~W.~Hawking and M.~J.~Perry,
 Nucl. Phys.
{\bf B318} (1978) 141.}

\lref\brih{D.~Brill and G.~T.~Horowitz, Phys. Lett. {\bf B262}
(1991)
437.}

\lref\gids{S.~B.~Giddings and A.~Strominger, Nucl. Phys.
{\bf B306}
(1988) 890.}

\lref\gidsone{S.~B.~Giddings and A.~Strominger, Phys. Lett.
{\bf B230}
(1989) 46.}

\lref\canhsw{P.~Candelas, G.~T.~Horowitz, A.~Strominger and
E.~Witten,
Nucl. Phys. {\bf B258} (1984) 46.}

\lref\bog{E.~B.~Bogomolnyi, Sov. J. Nucl. Phys. {\bf 24} (1976)
449.}

\lref\war{R.~S.~Ward, Comm. Math. Phys. {\bf 79} (1981) 317.}

\lref\warone{R.~S.~Ward, Comm. Math. Phys. {\bf 80} (1981) 563.}

\lref\wartwo{R.~S.~Ward, Phys. Lett. {\bf B158} (1985) 424.}

\lref\grop{D.~J.~Gross and M.~J.~Perry, Nucl. Phys. {\bf B226}
(1983)
29.}

\lref\ash{{\it New Perspectives in Canonical Gravity}, ed.
A.~Ashtekar,
(Bibliopolis, 1988).}

\lref\lic{A.~Lichnerowicz, {\it Th\' eories Relativistes de la
Gravitation et de l'Electro-magnetisme}, (Masson, Paris 1955).}

\lref\gol{H.~Goldstein, {\it Classical Mechanics},
Addison-Wesley,
1981.}

\lref\ros{P.~Rossi, Physics Reports, 86(6) 317-362.}

\lref\dixdp{J.~A.~Dixon, M.~J.~Duff and J.~C.~Plefka,
 Phys. Rev.
Lett.
{\bf 69} (1992) 3009.}

\lref\chad{J.~M.~Charap and M.~J.~Duff, Phys. Lett. {\bf B69}
(1977) 445.}

\lref\dufkexst{M.~J.~Duff and R.~R.~Khuri,
Nucl. Phys. {\bf B411} (1994) 473.}

\lref\khubifb{R.~R.~Khuri,
Phys. Rev. {\bf D48} (1993) 2947.}

\lref\khustab{R.~R.~Khuri, Phys. Lett. {\bf B307} (1993) 298.}

\lref\sor{R.~D.~Sorkin, Phys. Rev. Lett. {\bf 51} (1983) 87.}

\lref\dabh{A.~Dabholkar and J.~A.~Harvey,
 Phys. Rev. Lett. {\bf 63} (1989) 478.}

\lref\fels{A.~G.~Felce and T.~M.~Samols, Phys. Lett.
{\bf B308} (1993) 30.}

\lref\dufipss{M.~J.~Duff, T.~Inami, C.~N.~Pope, E.~Sezgin and
K.~S.~Stelle,
Nucl. Phys. {\bf B297}
(1988) 515.}

\lref\fujku{K.~Fujikawa and J.~Kubo,
 Nucl. Phys. {\bf B356} (1991) 208.}

\lref\cvet{M.~Cveti\v c, Phys. Rev. Lett. {\bf 71} (1993) 815.}

\lref\cvegs{M.~Cveti\v c, S. Griffies and H. H. Soleng, Phys.
 Rev. Lett. {\bf 71} (1993) 670; Phys. Rev. {\bf D48} (1993)
 2613.}

\lref\la{H. S. La, Phys. Lett. {\bf B315} (1993) 51.}

\lref\gresvy{B.~R.~Greene, A.~Shapere, C.~Vafa and S.~T.~Yau,
 Nucl. Phys.
{\bf B337} (1990) 1.}

\lref\fonilq{A.~Font, L.~Ib\'a\~nez, D.~Lust and F.~Quevedo,
Phys. Lett.
{\bf B249} (1990) 35.}

\lref\bin{P.~Bin\'etruy, Phys. Lett. {\bf B315} (1993) 80.}

\lref\koun{C.~Kounnas, in {\it Proceedings of INFN Eloisatron
Project, 26th Workshop: ``From Superstrings to Supergravity",
 Erice, Italy,
Dec. 5-12, 1992}, Eds. M.~Duff, S.~Ferrara and R.~Khuri,
(World Scientific, 1994).}

\lref\duftv{M.J. Duff, P.K. Townsend and P. van Nieuwenhuizen,
Phys. Lett. {\bf B122} (1983) 232.}

\lref\dufgt{M.J. Duff, G.W. Gibbons and P.K. Townsend
Phys. Lett. {\bf B} (1994) }

\lref\guv{R. G\"uven, Phys. Lett. {\bf B276} (1992) 49.}

\lref\guven{R. G\"uven, Phys. Lett. {\bf B212} (1988) 277.}

\lref\dobm{P.~Dobiasch and D.~Maison, Gen. Rel. Grav.
{\bf 14} (1982) 231.}

\lref\chod{A.~Chodos and S.~Detweiler, Gen. Rel. Grav.
{\bf 14} (1982) 879.}

\lref\pol{D.~Pollard, J. Phys. {\bf A16} (1983) 565.}

\lref\duffkr{M.~J.~Duff, S.~Ferrara, R.~R.~Khuri and J.~Rahmfeld,
 hep-th/9506057.}

\lref\lu{J. X. Lu,
 Phys. Lett. {\bf B313} (1993) 29.}

\lref\grel{R. Gregory and R. Laflamme, Phys. Rev. Lett.
{\bf 70} (1993) 2837.}

\lref\rom{L. Romans, Nucl. Phys. {\bf B276} (1986) 71.}

\lref\sala {A. Salam and E. Sezgin, {\it Supergravities in
Diverse Dimensions},
(North Holland/World Scientific, 1989).}

\lref\strath {J. Strathdee, Int. J. Mod. Phys. {\bf A2}
(1987) 273.}

\lref\dufliib{M.~J.~Duff and J.~X.~Lu,
 Nucl. Phys. {\bf B390} (1993) 276.}

\lref\nictv{H. Nicolai, P. K. Townsend and P. van
Nieuwenhuizen, Lett. Nuovo
Cimento {\bf 30} (1981) 315.}

\lref\towspan{P. K. Townsend, in Proceedings of the 13th GIFT
Seminar
on Theoretical Physics: {\it Recent Problems in Mathematical
Physics} Salamanca, Spain, 15-27 June, 1992.}

\lref\dufm{M.~J.~Duff and R.~Minasian,
 Nucl. Phys. {\bf B436} (1995) 507.}

\lref\gropy{D. J. Gross, R. D. Pisarski and L. G. Yaffe,
Rev. Mod. Phys.
{\bf 53} (1981) 43.}

\lref\rohw{R. Rohm and E. Witten,
 Ann. Phys. {\bf 170} (1986) 454.}

\lref\banddf{T. Banks, M. Dine, H. Dijkstra and W. Fischler,
Phys. Lett. {\bf B212} (1988) 45.}

\lref\ferkp{S. Ferrara, C. Kounnas and M. Porrati, Phys. Lett.
{\bf B181} (1986) 263.}

\lref\ter{M. Terentev, Sov. J. Nucl. Phys. {\bf 49} (1989) 713.}

\lref\hass{S. F. Hassan and A. Sen, Nucl. Phys. {\bf B375}
(1992) 103.}

\lref\mahs{J. Maharana and J. Schwarz, Nucl. Phys. {\bf B390}
(1993) 3.}

\lref\senrev{A. Sen,
 Int. J. Mod. Phys. {\bf A9} (1994) 3707.}

\lref\senone{A.~Sen,
Nucl. Phys. {\bf B404} (1993) 109.}

\lref\sentwo{A.~Sen,
  Int. J. Mod. Phys. {\bf A8} (1993) 5079.}

\lref\schsen{J.~H.~Schwarz and A.~Sen,
Nucl. Phys. {\bf B411} (1994) 35.}

\lref\schsentwo{J.~H.~Schwarz and A.~Sen,
 Phys. Lett. {\bf B312} (1993) 105.}

\lref\schtwo{J.~Schwarz,
CALT-68-1815.}

\lref\senph{A.~Sen, Phys. Lett. {\bf B303} (1993) 22.}

\lref\schwarz{J.~H.~Schwarz, CALT-68-1879, hepth/9307121.}

\lref\dufldr{M. J. Duff and J. X. Lu, Nucl. Phys. {\bf B347}
(1990) 394.}

\lref\salstr{A. Salam and J. Strathdee, Phys. Lett. {\bf B61}
(1976) 375.}

\lref\jjj{J. Gauntlett, J. Harvey and J. T. Liu,
Nucl. Phys. {\bf B409} (1993) 363.}

\lref\dupo{M.~J.~Duff and C.~N.~Pope, Nucl. Phys {\bf B255}
(1985) 355.}

\lref\sg{E.~Cremmer, S.~Ferrara, L.~Girardello and
 A.~Van Proeyen,
 Phys. Lett. {\bf B116} (1982) 231.}

\lref\sgone{E.~Cremmer, S.~Ferrara, L.~Girardello and
 A.~Van Proeyen,
 Nucl. Phys. {\bf B212} (1983) 413.}

\lref\fkp{S. Ferrara, C. Kounnas and M. Porrati,
Phys. Lett. {\bf B181} (1986) 263.}

\lref\fp{S. Ferrara and M. Porrati,
Phys. Lett. {\bf B216} (1989) 289.}

\lref\ghs{D.~Garfinkle, G.~T.~Horowitz and A.~Strominger,
Phys. Rev. {\bf D43}
(1991) 3140.}

\lref\hor{G.~T.~Horowitz, in Proceedings of Trieste '92,
{\it String theory and quantum gravity '92} p.55.}

\lref\gidps{S.~B.~Giddings, J.~Polchinski and A.~Strominger,
Phys. Rev. {\bf D48} (1993) 5784.}

\lref\shatw{A.~Shapere, S.~Trivedi and F.~Wilczek,
 Mod. Phys. Lett. {\bf A6}
(1991) 2677.}

\lref\klopv{R.~Kallosh, A.~Linde, T.~Ortin, A.~Peet and
A.~Van~Proeyen,
       Phys. Rev. {\bf D46} (1992) 5278.}

\lref\kal{R.~Kallosh, Phys. Lett. {\bf B282} (1992) 80.}

\lref\ko{R.~Kallosh and T.~Ortin, Phys. Rev. {\bf D48} (1993)
742.}

\lref\hw{C.~F.~E.~Holzhey and F.~Wilczek, Nucl. Phys.
{\bf B360} (1992) 447.}

\lref\dauria{R.~D'Auria, S.~Ferrara and M.~Villasante,
Class. Quant. Grav. {\bf 11} (1994) 481.}

\lref\gibp{G.~W.~Gibbons and M.~J.~Perry, Nucl. Phys.
{\bf B248} (1984) 629.}

\lref\salam{S. W. Hawking, Monthly Notices Roy. Astron. Soc.
 {\bf 152} (1971) 75; Abdus Salam in
   {\it Quantum Gravity: an Oxford Symposium} (Eds. Isham,
Penrose
and Sciama, O.U.P. 1975); G. 't Hooft, Nucl. Phys. {\bf B335}
(1990) 138.}

\lref\susskind{ L. Susskind, RU-93-44, hepth/9309145;
   J. G. Russo and L. Susskind, UTTG-9-94,
hepth/9405117.}

\lref\gibbons{ G. W. Gibbons, in {\it Supersymmetry,
Supergravity
and Related Topics}, Eds. F. del Aguila, J. A. Azcarraga and
L. E. Ibanez
(World Scientific, 1985).}

\lref\aichelburg{ P. Aichelburg and F. Embacher, Phys. Rev.
{\bf D37}
(1986) 3006.}

\lref\geroch{ R. Geroch, J. Math. Phys. {\bf 13} (1972) 394.}

\lref\hosoya{ A. Hosoya, K.
Ishikawa, Y. Ohkuwa and K. Yamagishi, Phys. Lett. {\bf B134}
(1984) 44.}

\lref\gibw{ G. W. Gibbons
and D. L. Wiltshire, Ann. of Phys. {\bf 167} (1986) 201.}

\lref\senprl{ A. Sen, Phys. Rev. Lett. {\bf 69}
(1992) 1006.}

\lref\schild{ G. C. Debney, R. P. Kerr and
   A. Schild, J. Math. Phys. {\bf 10} (1969) 1842.}

\lref\hort{G. T. Horowitz and A. A. Tseytlin, Phys. Rev.
{\bf D50} (1994) 5204.}

\lref\hortsey{G. T. Horowitz and A. A. Tseytlin,
Imperial/TP/93-94/51, UCSBTH-94-24, hepth/9408040;
Imperial/TP/93-94/54, UCSBTH-94-31, hepth/9409021.}

\lref\cvey{M. Cveti\v c and D. Youm, UPR-623-T, hepth/9409119.}

\lref\tseytlin{A. A. Tseytlin, Imperial-TP-93-94-46,
hepth/9407099.}

\lref\klim{C. Klimcik and A. A. Tseytlin, Nucl. Phys.
{\bf B424} (1994) 71.}

\lref\vafw{C. Vafa and E. Witten, hepth/9408074.}

\lref\glasgow{M. J. Duff, hepth/9410210.}

\lref\fere{R. C. Ferrell and D. M. Eardley,
Phys. Rev. Lett. {\bf 59} (1987) 1617.}

\lref\reyt{S. J. Rey and T. R. Taylor, Phys. Rev. Lett. {\bf 71}
(1993) 1132.}

\lref\senzwione{A. Sen and B. Zwiebach, Nucl. Phys.
{\bf B414} (1994) 649.}

\lref\senzwitwo{A. Sen and B. Zwiebach, Nucl. Phys.
{\bf B423} (1994) 580.}

\lref\bddo{T. Banks, A. Dabholkar, M. R. Douglas and
M. O'Loughlin, Phys. Rev. {\bf D45} (1992) 3607.}

\lref\bos{T. Banks, M. O'Loughlin and A. Strominger,
Phys. Rev. {\bf D47} (1993) 4476.}

\lref\kir{E. Kiritsis, Nucl. Phys. {\bf B405} (1993) 109.}

\lref\bakone{I. Bakas, Nucl. Phys. {\bf B428} (1994) 374.}

\lref\baktwo{I. Bakas, hepth/9410104.}

\lref\back{C. Bachas and E. Kiritsis, Phys. Lett. {\bf B325}
(1994) 103.}

\lref\bko{E. Bergshoeff, R. Kallosh and T. Ortin, hepth/9410230.}

\lref\bfrm{M. Bianchi, F. Fucito, G. C. Rossi and M. Martellini,
hepth/9409037.}

\lref\hult{C. M. Hull and P. K. Townsend, hepth/9410167.}

\lref\dufkmr{M. J. Duff, R. R. Khuri, R. Minasian and
J. Rahmfeld, Nucl. Phys. {\bf B418} (1994) 195.}

\lref\sentd{A. Sen, hepth/9408083.}

\lref\gresss{M. B. Green and J. Schwarz, Phys. Lett. {\bf B136}
(1984) 367.}

\lref\sie{W. Siegel, Phys. Lett. {\bf B128} (1983) 397.}

\lref\dir {P. A. M. Dirac, Pro. R. Soc. {\bf A133} (1931) 60.}

\lref\tho {G. t'Hooft, Nucl. Phys. {\bf B79} (1974) 276.}

\lref\pol {A. M. Polyakov, Sov. Phys. JETP Lett. {\bf 20}
(1974) 194.}

\lref\mono {C. Montonen and D. Olive, Phys. Lett. {\bf B72}
(1977) 117.}

\lref\col {S. Coleman, Phys. Rev. {\bf D11} (1975) 2088.}

\lref\grohmr {D. J. Gross, J. A. Harvey, E. Martinec and
 R. Rohm,
Nucl. Phys. {\bf B256} (1985) 253.}

\lref\ginone {P. Ginsparg, preprint LA--UR--91--9999 (1991),
 {\it Conformal Field Theory} Lectures given
at Trieste Summer School, Trieste, Italy, 1991.}

\lref\calhstwo {C. Callan, J. Harvey and A. Strominger,
{\it Supersymmetric String Solitons}, Lectures given at
Trieste Summer school, Trieste, Italy, 1991.}

\lref\sch {J. H. Schwarz, 1985, {\it Supersymmetry and Its
 Applications}
ed G. W. Gibbons {\it et al} (Cambridge University Press,
Cambridge) (1986).}

\lref\huglp {J. Hughes, J. Liu and J. Polchinski, Phys. Lett.
{\bf B180} (1986)
370.}

\lref\berst {E. Bergshoeff, E. Sezgin and P. Townsend,
 Phys. Lett. {\bf B189} (1987) 75.}

\lref\achetw {A. Achucarro, J. Evans, P. Townsend and
D. Wiltshire,
 Phys.
Lett. {\bf B198} (1987) 441.}

\lref\belpst {A. A. Belavin, A. M. Polyakov, A. S. Schwartz and
Yu. S. Tyupkin,
Phys. Lett. {\bf B59} (1975) 85.}

\lref\oset {D. O'Se and D. H. Tchrakian, Lett. Math. Phys.
{\bf 13} (1987) 211.}

\lref\groks {B. Grossman, T. W. Kephart and J. D. Stasheff,
 Commun. Math. Phys. {\bf 96} (1984) 431;
Commun. Math. Phys. {\bf 100} (1985) 311.}

\lref\grokstwo {B. Grossman, T. W. Kephart and J. D. Stasheff,
 Phys. Lett. {\bf B220}
 (1989)  431}

\lref\tch{D. H. Tchrakian, Phys. Lett. {\bf B150}  (1985)  360.}

\lref\fubn{S. Fubini and H. Nicolai,
 Phys. Lett. {\bf B155}  (1985)  369}

\lref\fain{D. B. Fairlie and J. Nuyts,
 J. Phys. {\bf A17}  (1984)  2867.}

\lref\str {A. Strominger,  Nucl. Phys. {\bf B343}
(1990) 167.}

\lref\duflhs {M. J. Duff and J. X. Lu,
Phys. Rev. Lett. {\bf 66} (1991) 1402.}

\lref\godo {P. Goddard and D. Olive, Rep. Prog. Phys. {\bf 41}
(1978) 1357.}

\lref\colone {S. Coleman, Pro. 1975 Int. School on Subnuclear
Physics,
Erice, ed A. Zichichi  (Plenum, New York)   (1977); Pro. 1981
Int. School
on Subnuclear Physics, Erice, ed A. Zichichi (Plenum, New York)
(1983).}

\lref\wuy {T. T. Wu and C. N. Yang, Nucl. Phys. {\bf B107}
(1976) 365.}

\lref\dufldl {M. J. Duff and J. X. Lu, Class. Quantum Grav.
{\bf 9} (1992) 1.}

\lref\tei {C. Teitelboim,
Phys. Lett. {\bf B167} (1986) 69.}

\lref\nep {R. I. Nepomechie,
Phys. Rev. {\bf D31} (1984) 1921.}

\lref\raj {R. Rajaraman, Solitons and Instantons
(North--Holland, Amsterdam)
(1982).}

\lref\wito {E. Witten and D. Olive, Phys. Lett.
{\bf B78} (1978) 97.}

\lref\pras {M. K. Prasad and C. M. Sommerfield, Phys. Rev. Lett.
{\bf 35} (1975) 760.}

\lref\corg {E. Corrigan and P. Goddard, Commun. Math. Phys.
{\bf 80} (1981)
575.}

\lref\godno {P. Goddard, J. Nuyts and D. Olive, Nucl. Phys.
{\bf B125} (1977)
1.}

\lref\osb {H. Osborn, Phys. Lett. {\bf B83} (1979) 321.}

\lref\egugh {T. Eguchi, P. B. Gilkey and A. J. Hanson,
Phys. Rep. {\bf 66}
(1980) 213.}

\lref\corf {E. F. Corrigan and D. B. Fairlie, Phys. Lett.
{\bf B67} (1977) 69.
}

\lref\atidhm {M. F. Atiyah, V. G. Drinfeld, N. J. Hitchin and
Y. I. Manin,
Phys. Lett. {\bf A65} (1978) 185.}

\lref\dun {A. R. Dundarer,
 Mod. Phys. Lett. {\bf A5} (1991) 409.}

\lref\cha {A. H. Chamseddine, Phys. Rev. {\bf D24} (1981) 3065.}

\lref\berrwv {E. A. Bergshoeff, M. de Roo, B. de Wit and
 P. van Nieuwenhuizen,
Nucl. Phys. {\bf B195}  (1982)  97}

\lref\cham{G. F. Chapline and N. S. Manton, Phys. Lett.
{\bf B120} (1983) 105.}

\lref\gatn {S. J. Gates and H. Nishino, Phys. Lett. {\bf B173}
(1986) 52.}

\lref\sala{A. Salam and E. Sezgin, Physica Scripta {\bf 32}
(1985) 283.}

\lref\duf {M. J. Duff,  Class.
 Quantum  Grav. {\bf 5} (1988) 189.}

\lref\duflfb {M. J. Duff and J. X. Lu, Nucl. Phys. {\bf B354}
(1991) 141.}

\lref\dabghr {A. Dabholkar, G. W. Gibbons, J. A. Harvey and
F. Ruiz Ruiz,
 Nucl. Phys. {\bf B340} (1990) 33.}

\lref\berdps {E. Bergshoeff, M. J. Duff, C. N. Pope and
E. Sezgin,
 Phys. Lett.
{\bf B199} (1987) 69.}

\lref\tow {P. Townsend,
Phys. Lett. {\bf B202} (1988) 53.}

\lref\dufhis {M. J. Duff, P. S. Howe, T. Inami and K. Stelle,
 Phys. Lett.
{\bf B191} (1987) 70.}

\lref\dufs {M. J. Duff and K. Stelle, Phys. Lett. {\bf B253}
(1991) 113.}

\lref\berst {E. Bergshoeff, E. Sezgin and P. Townsend, Ann.
 Phys. {\bf 199}
(1990) 340.}

\lref\duflrsfd {M. J. Duff and J. X. Lu, Nucl. Phys. {\bf B354}
(1991) 129.}

\lref\gresone {M. Green and J. Schwarz, Phys. Lett. {\bf B151}
(1985) 21.}

\lref\bercgw {C. W. Bernard, N. H. Christ, A. H. Guth and
E. J. Weinberg,
 Phys. Rev.
 {\bf D16} (1977) 2967.}

\lref\hars {J. Harvey and A. Strominger,
 Phys. Rev. Lett. {\bf 66} (1991) 549.}

\lref\gres {M. Green and J. Schwarz, Phys. Lett. {\bf B149}
(1984) 117.}

\lref\elljm {J. Ellis, P. Jetzer and L. Mizrachi,
Nucl. Phys. {\bf B303} (1988)
1.}

\lref\dixds {J. Dixon, M. J. Duff and E. Sezgin, preprint
CTP--TAMU--101/91
 (1991).}

\lref\berrs {E. Bergsheoff, M. Rakowski and E. Sezgin, Phys.
 Lett. {\bf
B185}  (1987)  371}

\lref\berd{E. Bergsheoff and M. de Roo, Nucl. Phys. {\bf B328}
(1989)
439}

\lref\dersw{M. de Roo, H. Suelmann and A. Wiedemann, preprint
UG--1/92  (1992).}

\lref\gresw {M. Green, J. Schwarz and E. Witten, Superstring
theory, Cambridge University Press, Cambridge, 1987.}

\lref\duflloop {M. J. Duff and J. X. Lu, Nucl. Phys. {\bf B357}
  (1991)  534.}

\lref\ven {G. Veneziano,
Europhys. Lett. {\bf 2}  (1986)  199.}

\lref\cain {Y. Cai and C. A. Nunez, Nucl. Phys. {\bf B287}
(1987)  41}

\lref\gros{D. J. Gross and J. Sloan, Nucl. Phys. {\bf B291}
(1987)  41.}

\lref\ellm {J. Ellis and L. Mizrachi,
  Nucl. Phys. {\bf B327}  (1989)  595.}

\lref\calfmp {C. G. Callan, D. Friedan, E. J. Martinec and
M. J. Perry,
Nucl. Phys. {\bf B262}  (1985)  593.}

\lref\grestwo {M. Green and J. Schwarz, Phys. Lett. {\bf B173}
 (1986)  52.}

\lref\lin {U. Lindstrom, in Supermembranes and Physics in 2 + 1
Dimensions, ed. M. J. Duff, C. N. Pope and E. Sezgin
(World Scientific,
Singapore) (1990).}

\lref\dufone {M. J. Duff, Class. Quantum Grav. {\bf 6}  (1989)
1577.}

\lref\callny {C. Callan, C. Lovelace, C. Nappi and S. Yost,
Nucl. Phys.
{\bf B308}  (1988)  221.}

\lref\frat {E. Fradkin and A. Tseytlin, Phys. Lett. {\bf B158}
(1985)  316.}

\lref\duflselft {M. J. Duff and J. X. Lu, Phys. Lett.
{\bf B273}  (1991)  409.}

\lref\hors {G. Horowitz and A. Strominger, Nucl. Phys.
{\bf B360}  (1991) 197. }

\lref\witone {E. Witten,  Phys. Lett. {\bf B86}  (1979)
283.}

\lref\schone {J. Schwarz, Nucl. Phys. {\bf B226}  (1983)  269.}

\lref\zwa {D. Zwanziger, Phys. Rev. {\bf 176}  (1968)  1480,
1489.}

\lref\schwing {J. Schwinger, Phys. Rev. {\bf 144} (1966) 1087;
{\bf 173}
(1968) 1536.}

\lref\gibt{G.W. Gibbons and P.K. Townsend, Phys. Rev. Lett.
{\bf 71} (1993)
3754.}

\lref\duflblacks {M. J. Duff and J. X. Lu,
 Nucl. Phys. {\bf B416} (1994) 301.}

\lref\dufklsin {M. J. Duff, R. R. Khuri and J. X. Lu,
 Nucl. Phys. {\bf B377}
(1992) 281.}

\lref\calk {C.~G.~Callan and R.~ R.~Khuri,
Phys. Lett. {\bf B261} (1991) 363.}

\lref\gib {G. W. Gibbons, Nucl. Phys. {\bf B207} (1982) 337.}

\lref\gibm {G. W. Gibbons and K. Maeda, Nucl. Phys. {\bf B298}
(1988) 741.}

\lref\rey{S. J. Rey, in Proceedings of Tuscaloosa
Workshop on Particle Physics, Tuscaloosa, Alabama, 1989.}

\lref\reyone {S. J.~Rey, Phys. Rev. {\bf D43} (1991) 526.}

\lref\antben {I.~Antoniadis, C.~Bachas, J.~Ellis and
 D.~V.~Nanopoulos,
Phys. Lett. {\bf B211} (1988) 393.}

\lref\antbenone {I.~Antoniadis, C.~Bachas, J.~Ellis and
D.~V.~Nanopoulos,
Nucl. Phys. {\bf B328} (1989) 117.}

\lref\mett {R.~R.~Metsaev and A.~A.~Tseytlin, Phys. Lett.
{\bf B191} (1987) 354.}

\lref\mettone {R.~R.~Metsaev and A.~A.~Tseytlin,
Nucl. Phys. {\bf B293} (1987) 385.}

\lref\calkp {C.~G.~Callan,
I.~R.~Klebanov and M.~J.~Perry, Nucl. Phys. {\bf B278} (1986)
78.}

\lref\lov {C.~Lovelace, Phys. Lett. {\bf B135} (1984) 75.}

\lref\friv {B.~E.~Fridling and A.~E.~M.~Van de Ven,
Nucl. Phys. {\bf B268} (1986) 719.}

\lref\gepw {D.~Gepner and E.~Witten, Nucl. Phys. {\bf B278}
(1986) 493.}

\lref\din {M.~Dine, Lectures delivered at
TASI 1988, Brown University (1988) 653.}

\lref\berdone {E.~A.~Bergshoeff and M.~de Roo, Phys. Lett.
{\bf B218} (1989)
210.}

\lref\calhs{C.~G.~Callan, J.~A.~Harvey and A.~Strominger,
Nucl. Phys.
{\bf B359} (1991) 611.}

\lref\calhsone{C.~G.~Callan, J.~A.~Harvey and A.~Strominger,
Nucl. Phys.
{\bf B367} (1991) 60.}

\lref\thoone{G.~'t~Hooft, Phys. Rev. Lett., {\bf 37} (1976) 8.}

\lref\wil{F.~Wilczek, in
{\it Quark confinement and field theory},
Ed. D.~Stump and D.~Weingarten, John Wiley and Sons, New York
(1977).}

\lref\jacnr{R.~Jackiw, C.~Nohl and C.~Rebbi, Phys. Rev.
{\bf D15} (1977)
1642.}

\lref\khuinst{R.~R.~Khuri, Phys. Lett.
{\bf B259} (1991) 261.}

\lref\khumant{R.~R.~Khuri, Nucl. Phys.
 {\bf B376} (1992) 350.}

\lref\khumono{R.~R.~Khuri,
 Phys. Lett. {\bf B294} (1992) 325.}

\lref\khumonscat{R.~R.~Khuri,
Phys. Lett. {\bf B294} (1992) 331.}

\lref\khumonex{R.~R.~Khuri,
Nucl. Phys. {\bf B387} (1992) 315.}

\lref\khumonin{R.~R.~Khuri,
 Phys. Rev. {\bf D46} (1992) 4526.}

\lref\khugeo{R.~R.~Khuri,
Phys. Lett. {\bf 307} (1993) 302.}

\lref\khuscat{R.~R.~Khuri,
 Nucl. Phys. {\bf B403} (1993) 335.}

\lref\khuwind {R.~R.~Khuri, Phys. Rev. {\bf D48} (1993) 2823.}

\lref\gin{P.~Ginsparg, Lectures delivered at
Les Houches summer session, June 28--August 5, 1988.}

\lref\alljj{R. W. Allen, I. Jack and D. R. T. Jones,
 Z. Phys. {\bf C41}
(1988) 323.}

\lref\sev{A. Sevrin, W. Troost and A. van Proeyen,
Phys. Lett. {\bf B208} (1988) 447.}

\lref\schout{K. Schoutens, Nucl. Phys. {\bf B295} [FS21] (1988)
634.}

\lref\harl{J.~A.~Harvey and J.~Liu, Phys. Lett. {\bf B268}
(1991) 40.}

\lref\man{N.~S.~Manton, Nucl. Phys. {\bf B126} (1977) 525.}

\lref\manone{N.~S.~Manton, Phys. Lett. {\bf B110} (1982) 54.}

\lref\mantwo{N.~S.~Manton, Phys. Lett. {\bf B154} (1985) 397.}

\lref\atihone{M.~F.~Atiyah and N.~J.~Hitchin, Phys. Lett.
{\bf A107}
(1985) 21.}

\lref\atihtwo{M.~F.~Atiyah and N.~J.~Hitchin, {\it The Geometry
and
Dynamics of Magnetic Monopoles}, Princeton University Press,
1988.}

\lref\polc{J.~Polchinski, Phys. Lett. {\bf B209} (1988) 252.}

\lref\gibhp{G.~W.~Gibbons and S.~W.~Hawking, Phys. Rev.
{\bf D15}
(1977) 2752.}

\lref\gibhpone{G.~W.~Gibbons, S.~W.~Hawking and M.~J.~Perry,
 Nucl. Phys.
{\bf B318} (1978) 141.}

\lref\brih{D.~Brill and G.~T.~Horowitz, Phys. Lett. {\bf B262}
(1991)
437.}

\lref\gids{S.~B.~Giddings and A.~Strominger, Nucl. Phys.
{\bf B306}
(1988) 890.}

\lref\gidsone{S.~B.~Giddings and A.~Strominger, Phys. Lett.
{\bf B230}
(1989) 46.}

\lref\canhsw{P.~Candelas, G.~T.~Horowitz, A.~Strominger and
E.~Witten,
Nucl. Phys. {\bf B258} (1984) 46.}

\lref\bog{E.~B.~Bogomolnyi, Sov. J. Nucl. Phys. {\bf 24} (1976)
449.}

\lref\war{R.~S.~Ward, Comm. Math. Phys. {\bf 79} (1981) 317.}

\lref\warone{R.~S.~Ward, Comm. Math. Phys. {\bf 80} (1981) 563.}

\lref\wartwo{R.~S.~Ward, Phys. Lett. {\bf B158} (1985) 424.}

\lref\grop{D.~J.~Gross and M.~J.~Perry, Nucl. Phys. {\bf B226}
(1983)
29.}

\lref\ash{{\it New Perspectives in Canonical Gravity}, ed.
A.~Ashtekar,
Bibliopolis, 1988.}

\lref\lic{A.~Lichnerowicz, {\it Th\' eories Relativistes de la
Gravitation et de l'Electro-magnetisme}, (Masson, Paris 1955).}

\lref\gol{H.~Goldstein, {\it Classical Mechanics},
Addison-Wesley,
1981.}

\lref\ros{P.~Rossi, Physics Reports, 86(6) 317-362.}

\lref\dixdp{J.~A.~Dixon, M.~J.~Duff and J.~C.~Plefka,
{\it Putting String/Fivebrane Duality to the Test} Phys. Rev.
Lett.
{\bf 69} (1992) 3009.}

\lref\chad{J.~M.~Charap and M.~J.~Duff, Phys. Lett. {\bf B69}
(1977) 445.}

\lref\dufkexst{M.~J.~Duff and R.~R.~Khuri,
Nucl. Phys. {\bf B411} (1994) 473.}

\lref\khubifb{R.~R.~Khuri,
Phys. Rev. {\bf D48} (1993) 2947.}

\lref\sor{R.~D.~Sorkin, Phys. Rev. Lett {\bf 51} (1983) 87.}

\lref\dabh{A.~Dabholkar and J.~A.~Harvey,
 Phys. Rev. Lett. {\bf 63} (1989) 478.}

\lref\fels{A.~G.~Felce and T.~M.~Samols, Phys. Lett.
{\bf B308} (1993) 30.}

\lref\dufipss{M.~J.~Duff, T.~Inami, C.~N.~Pope, E.~Sezgin and
K.~S.~Stelle,
Nucl. Phys. {\bf B297}
(1988) 515.}

\lref\fujku{K.~Fujikawa and J.~Kubo,
 Nucl. Phys. {\bf B356} (1991) 208.}

\lref\cvet{M.~Cveti\v c, Phys. Rev. Lett. {\bf 71} (1993) 815.}

\lref\gresvy{B.~R.~Greene, A.~Shapere, C.~Vafa and S.~T.~Yau,
 Nucl. Phys.
{\bf B337} (1990) 1.}

\lref\fonilq{A.~Font, L.~Ib\'a\~nez, D.~Lust and F.~Quevedo,
Phys. Lett.
{\bf B249} (1990) 35.}

\lref\bin{P.~Bin\'etruy, Phys. Lett. {\bf B315} (1993) 80.}

\lref\koun{C.~Kounnas, in {\it Proceedings of INFN Eloisatron
Project, 26th Workshop: ``From Superstrings to Supergravity",
 Erice, Italy,
Dec. 5-12, 1992}, Ed. M.~Duff, S.~Ferrara and R.~Khuri,
World Scientific.}

\lref\gibht{G.W. Gibbons, G. Horowitz and P.K. Townsend, in
preparation.}

\lref\duftv{M.J. Duff, P.K. Townsend and P. van Nieuwenhuizen,
Phys. Lett. {\bf B122} (1983) 232.}

\lref\dufgt{M.J. Duff, G.W. Gibbons and P.K. Townsend
Phys. Lett. {\bf B} (1994) }

\lref\guv{R. Guven, Phys. Lett. {\bf B276} (1992) 49.}

\lref\dobm{P.~Dobiasch and D.~Maison, Gen. Rel. Grav.
{\bf 14} (1982) 231.}

\lref\chod{A.~Chodos and S.~Detweiler, Gen. Rel. Grav.
{\bf 14} (1982) 879.}

\lref\pol{D.~Pollard, J. Phys. {\bf A16} (1983) 565.}

\lref\duffkr{M.~J.~Duff, S.~Ferrara, R.~R.~Khuri and J.~Rahmfeld,
 in preparation.}

\lref\lu{J. X. Lu,
 Phys. Lett. {\bf B313} (1993) 29.}

\lref\grel{R. Gregory and R. Laflamme, ``Black strings
and $p$-branes
are unstable", EFI-93-02.}

\lref\rom{L. Romans, Nucl. Phys. {\bf B276} (1986) 71.}

\lref\sala {A. Salam and E. Sezgin, ``Supergravities in
Diverse Dimensions'',
(North Holland/World Scientific 1989).}

\lref\strath {J. Strathdee, Int. J. Mod. Phys. {\bf A2}
(1987) 273.}

\lref\dufliib{M.~J.~Duff and J.~X.~Lu,
 Nucl. Phys.{\bf B390} (1993) 276.}

\lref\nictv{H. Nicolai, P. K. Townsend and P. van Nieuwenhuizen, Lett. Nuovo
Cimento {\bf 30} (1981) 315.}

\lref\towspan{P. Townsend, Spanish lectures.}

\lref\dufm{M.~J.~Duff and R.~Minasian, Nucl. Phys.
{\bf B436} (1995) 507.}

\lref\gropy{D. J. Gross, R. D. Pisarski and L. G. Yaffe,
Rev. Mod. Phys.
{\bf 53} (1981) 43.}

\lref\rohw{R. Rohm and E. Witten,
 Ann. Phys. {\bf 170} (1986) 454.}

\lref\banddf{T. Banks, M. Dine, H. Dijkstra and W. Fischler,
Phys. Lett. {\bf B212} )1988) 45.}

\lref\ferkp{S. Ferrara, C. Kounnas and M. Porrati, Phys. Lett.
{\bf B181} (1986) 263.}

\lref\ter{M. Terentev, Sov. J. Nucl. Phys. {\bf 49} (1989) 713.}

\lref\hass{S. F. Hassan and A. Sen, Nucl. Phys. {\bf B375}
(1992) 103.}

\lref\mahs{J. Maharan and J. Schwarz, Nucl. Phys. {\bf B390}
(1993) 3.}

\lref\senrev{A. Sen,
 Int. J. Mod. Phys. {\bf A9} (1994) 3707.}

\lref\senone{A.~Sen,
Nucl. Phys. {\bf B404} (1993) 109.}

\lref\sentwo{A.~Sen,
  Int. J. Mod. Phys. {\bf A8} (1993) 5079.}

\lref\schsen{J.~H.~Schwarz and A.~Sen,
Nucl. Phys. {\bf B411} (1994) 35.}

\lref\schsentwo{J.~H.~Schwarz and A.~Sen,
 Phys. Lett. {\bf B312} (1993) 105.}

\lref\schtwo{J.~Schwarz,
CALT-68-1815.}

\lref\senph{A.~Sen, Phys. Lett. {\bf B303} (1993) 22.}

\lref\schwarz{J.~H.~Schwarz, CALT-68-1879 (hep-th@9307121).}

\lref\dufldr{M. J. Duff and J. X. Lu, Nucl. Phys. {\bf B354}
(1994) 473.}

\lref\dirszw{P. Dirac, Proc. R. Soc. {\bf A133} (1931) 60;
J. Schwinger, Phys. Rev. {\bf 144} (1966) 1087; {\bf 173} (1968) 1536;
D. Zwanziger, Phys. Rev. {\bf 176} (1968) 1480, 1489;
E. Witten, Phys. Lett. {\bf B86} (1979) 283.}

\lref\salstr{A. Salam and J. Strathdee, Phys. Lett. {\bf B61}
(1976) 375.}

\lref\jjj{J. Gauntlett, J. Harvey and J. T. Liu.}

\lref\dupo{M.~J.~Duff and C.~N.~Pope, Nucl. Phys {\bf B255}
(1985) 355.}

\lref\sgone{E.~Cremmer, S.~Ferrara, L.~Girardello,
 A.~Van Proeyen,
 Nucl. Phys. {\bf B212} (1983) 413.}

\lref\ghs{D.~Garfinkle, G.~T.~Horowitz and A.~Strominger,
Phys. Rev. {\bf D43}
(1991) 3140.}

\lref\hor{G.~T.~Horowitz, {\it The Dark Side of
String Theory: Black Strings and Black Holes},
in Proceedings of Trieste Summer School, Trieste, Italy, 1992,
 p.55.}

\lref\gidps{S.~B.~Giddings, J.~Polchinski and A.~Strominger,
Phys. Rev. {\bf D48} (1993) 5784.}

\lref\shatw{A.~Shapere, S.~Trivedi and F.~Wilczek,
 Mod. Phys. Lett. {\bf A6}
(1991) 2677.}

\lref\klopv{R.~Kallosh, A.~Linde, T.~Ortin, A.~Peet and
A.~Van~Proeyen,
       Phys. Rev. {\bf D46} (1992) 5278.}

\lref\kal{R.~Kallosh, Phys. Lett. {\bf B282} (1992) 80.}

\lref\ko{R.~Kallosh and T.~Ortin, Phys. Rev. {\bf D48} (1993)
742.}

\lref\hw{C.~F.~E.~Holzhey and F.~Wilczek, Nucl. Phys.
{\bf B360} (1992) 447.}

\lref\dauria{R.~D'Auria, S.~Ferrara and M.~Villasante,
CERN-TH 6914/93,
      POLFIS-TH 04/93, UCLA/93/TEP/18 (hep-th@9306125).}

\lref\gibp{G.~W.~Gibbons and M.~J.~Perry, Nucl. Phys.
{\bf B248} (1984) 629.}

\lref\salam{S. W. Hawking, Monthly Notices Roy. Astron. Soc.
 {\bf 152} (1971) 75; Abdus Salam in
   {\it Quantum Gravity: an Oxford Symposium} (Eds. Isham,
Penrose
and Sciama, O.U.P. 1975); G. 't Hooft, Nucl. Phys. {\bf B335}
(1990) 138.}

\lref\susskind{ L. Susskind, RU-93-44, hep-th/9309145;
   J. G. Russo and L. Susskind, UTTG-9-94,
hep-th/9405117.}

\lref\gibbons{ G. W. Gibbons, in {\it Supersymmetry,
Supergravity
and Related Topics}, eds. F. del Aguila, J. A. Azcarraga and
L. E. Ibanez
(World Scientific, 1985).}

\lref\aichelburg{ P. Aichelburg and F. Embacher, Phys. Rev.
{\bf D37}
(1986) 3006.}

\lref\geroch{ R. Geroch, J. Math. Phys. {\bf 13} (1972) 394.}

\lref\hosoya{ A. Hosoya, K.
Ishikawa, Y. Ohkuwa and K. Yamagishi, Phys. Lett. {\bf B134}
(1984) 44.}

\lref\gibw{ G. W. Gibbons
and D. L. Wiltshire, Ann. of Phys. {\bf 167} (1986) 201.}

\lref\senprl{ A. Sen, Phys. Rev. Lett. {\bf 69}
(1992) 1006.}

\lref\schild{ G. C. Debney, R. P. Kerr and
   A. Schild, J. Math. Phys. {\bf 10} (1969) 1842.}

\lref\hort{G. T. Horowitz and A. A. Tseytlin,
 hepth/9409067.}

\lref\tseytlin{A. A. Tseytlin,
hepth/9407099.}

\lref\klim{C. Klimcik and A. A. Tseytlin,
{\it Exact four-dimensional
 string solutions and Toda-like sigma models from null-gauged
 WZNW models},
hepth/9402120.}

\newsec{Introduction}
In this review I discuss some basic results in the
study of classical solitonic and black hole solutions
of string theory. One motivation in this endeavour is that
the existence of these Plack-scale solitons may shed light on
the nature of string theory as a finite theory of quantum
gravity. Furthermore, there is the possibility of adapting
to string theory nonperturbative
methods from the physics of solitons and
instantons already employed in field theory. For example,
 the stringy analogs of Yang-Mills instantons may be used to
explore tunneling between string vacua and
thus lead to a better understanding of the nature of the vacuum
in string theory. Finally, these soliton and black hole solutions
point to interesting connections between the various spacetime
and worldsheet dualities in string theory.

I begin in section 2 with a description of the axionic
instanton solution in the gravitational sector of the string,
first discovered in \refs{\rey,\reyone}\ and which represents a
stringy analog
of the 't Hooft ansatz \refs{\thoone\wil\corf{--}\jacnr}. In
particular, the generalized curvature of the string solution,
with torsion coming from the antisymmetric tensor field strength
$H_3$
(and hence the name ``axionic instanton'', since in four
dimensions
$H_3$ is dual to an axion field)
obeys a (anti) self-duality condition identical to that obeyed
by the field strength of the Yang-Mills instanton
\refs{\khuinst,\khumonin}. In ten-dimensional heterotic string
theory the axionic instanton manifests
itself as a $5+1$-dimensional soliton solution, the so-called
``fivebrane'' \refs{\dufone,\str,\duflfb} and whose existence
was predicted by the string/fivebrane duality conjecture
\refs{\duf,\duflrsfd}. Related solutions with axionic instanton
structure are also briefly discussed \refs{\calhs,\khubifb}.

In section 3 I discuss toroidal compactifications of the
axionic
instanton/fivebrane to four dimensions and obtain supersymmetric
monopole \khumono, string and domain wall solutions \dufkexst\
which
break half the
spacetime supersymmetries of $N=4$, $D=4$ heterotic string
theory.

The string soliton solution is singled out in section 4, where
it is
observed to be the solitonic dual of the fundamental string
solution
of \dabghr\ in four-dimensions. Another way of saying this is
that
the string/fivebrane duality conjecture manifests itself as
(effective)
string/string duality in $D=4$ \dufkexst.
One attraction of this reduction is that
a dual string theory is probably far easier to construct than a
fundamental
fivebrane theory. More immediately, four-dimensional
string/string
duality is seen to interchange two other dualities in string
theory \refs{\schsen,\bin,\dufkexst}:
target space duality, already established
in various compactifications, and strong/weak coupling duality,
shown
in the low-energy limit but conjectured to be an exact symmetry
of
string theory.

The four-dimensional solitons represent extremal limits,
saturating a
Bogomol'nyi bound \bog\ between mass and charge, of
two-parameter families
of black hole solutions \dufkmr. These black hole
 generalizations,
as well as their connections with ten-dimensional and
four-dimensional
dualities, are discussed in section 5.

In section 6 I study the dynamics of the four-dimensional
solitons from
two different viewpoints \khuscat. The first involves computing
the
 Manton metric
on moduli space \mantwo, whose geodesics represent the motion
of quasi-static
solutions in the static solution manifold, and which represent a
low-velocity approximation to the actual dynamics of the
solitons. The
second approach calculates the four-point amplitude for the
 scattering
of winding string states, the nearest approximation in string
theory to
solitonic string states. Both computations yield trivial
scattering to
leading order in the velocities (i.e. zero-dynamical force to l
eading
order) in direct contrast to analogous computations for BPS
monopoles
\refs{\pras,\bog}.

In section 7 I present new string solutions corresponding to
more
intricate toroidal compactifications \duffkr. Interesting
connections
are
made between the number of preserved supersymmetries and the
 nature
of the target space duality group. The role of the axionic
instanton is
again seen to be crucial in this respect. Analogous solutions
are also
seen to arise in more realistic $N=1$ and $N=2$
compactifications.

Finally, in section 8 I discuss future directions in this
subfield of string theory and
suggest some open problems. Earlier reviews which deal
more extensively with the conformal field theoretic aspects
of soliton and black hole solutions of string theory may
be found in \refs{\calhstwo,\hor}.

\newsec{Axionic Instanton}

Consider the four-dimensional Euclidean action
\eqn\ymact{S=-{1\over 2g^2}\int d^4y {\rm tr} F_{mn}F^{mn},
\qquad\qquad m,n =1,2,3,4.}
For gauge group $SU(2)$, the fields may be written as $A_m=(g/2i)
\sigma^a A_m^a$ and $F_{mn}=(g/2i)\sigma^a F_{mn}^a$\ \
(where $\sigma^a$, $a=1,2,3$ are the $2\times 2$ Pauli matrices).
The equation of motion derived from this action is solved by the
't Hooft ansatz
\refs{\thoone\wil\corf{--}\jacnr}
\eqn\hfanstz{A_{mn}=i \overline{\Sigma}_{mn}\partial_n \ln f,}
where $\overline{\Sigma}_{mn}=\overline{\eta}^{imn}(\sigma^i/2)$
for $i=1,2,3$, where
\eqn\hfeta{\eqalign{\overline{\eta}^{imn}=-\overline{\eta}^{inm}
&=\epsilon^{imn},\qquad\qquad m,n=1,2,3,\cr
&=-\delta^{im},\qquad\qquad n=4 \cr}}
and where $f^{-1}\Box\ f=0$. The above solution obeys the
 self-duality
condition
\eqn\fsd{F_{mn}=\tilde F_{mn}={1\over 2}
\epsilon_{mn}{}^{kl} F_{kl}.}
The ansatz for the anti-self-dual solution
$F_{mn}=-\tilde F_{mn}=-{1\over 2} \epsilon_{mn}{}^{kl} F_{kl}$
is similar, with the $\delta$-term in \hfeta\ changing sign.
To obtain a multi-instanton solution, one solves for $f$ in the
four-dimensional space to obtain
\eqn\finst{f=1+\sum_{i=1}^k{\rho_i^2\over |\vec y - \vec a_i|^2},}
where $\rho_i$ is the instanton scale size, $\vec a_i$ the
location in
four-space of the $i$th instanton and
\eqn\inwind{k = {1\over 16 \pi^2} \int_{M^4} tr F^2}
is the instanton number. Note that this solution has $5k$
parameters,
while the most general (anti) self-dual solution has $8k$
parameters, or
$8k-3$ if one excludes the $3$ zero modes associated with global $SU(2)$
rotations.

Now consider the bosonic string sigma model action \lov
\eqn\sigmod{I={1\over 4\pi\alpha'}\int d^2x\sqrt{\gamma}
\left(\gamma^{ab}
\partial_aX^M\partial_bX^N g_{MN}+i\epsilon^{ab}\partial_aX^M
\partial_b
X^N B_{MN}+\alpha'R^{(2)}\phi\right),}
where $g_{MN}$ is the sigma model metric, $\phi$ is the dilaton
 and $B_{MN}$
is the antisymmetric tensor, and where $\gamma_{ab}$ is the
worldsheet metric
and $R^{(2)}$ the two-dimensional curvature. A classical solution
 of
bosonic string theory corresponds to Weyl invariance of \sigmod\
\mett.
It turns out that any dilaton function satisfying
$e^{-2\phi}\Box\ e^{2\phi}=0$ with
\eqn\bansatz{\eqalign{\met&=e^{2\phi}\delta_{mn}\qquad m,n=1,2,3,4,\cr
g_{\mu\nu}&=\eta_{\mu\nu}\qquad\quad   \mu,\nu=0,5,...,25,\cr
H_{mnp}&=\pm 2\epsilon_{mnpk}\partial^k\phi
\qquad m,n,p,k=1,2,3,4,\cr}}
where $H=dB$, is a tree-level solution of \sigmod.
The ansatz \bansatz\ in fact
possesses a (anti) self-dual structure in the subspace $(1234)$,
which can be seen as follows.
We define a generalized curvature $\grijkl$ in terms of the
standard
curvature $\rijkl$ and $H_{\mu\alpha\beta}$ \friv:
\eqn\gcurv{\grijkl=\rijkl+{1\over
2}\left(\nabla_lH^i{}_{jk}-\nabla_kH^i{}_{jl}\right)
+{1\over 4}\left(H^m{}_{jk}H^i{}_{lm}-H^m{}_{jl}H^i{}_{km}\right).}
One can also define $\grijkl$ as the Riemann tensor generated
by the generalized Christoffel symbols
$\hat\Gamma^\mu_{\alpha\beta}$
where  $\hat\Gamma^\mu_{\alpha\beta}=\Gamma^\mu_{\alpha\beta}
-(1/2) H^\mu{}_{\alpha\beta}$.
Then we can express the generalized curvature in covariant form
in terms of the dilaton field as \khuinst\
\eqn\gcurvphi{\grijkl=\delta_{il}\nabla_k\nabla_j\phi
-\delta_{ik}\nabla_l\nabla_j\phi+\delta_{jk}\nabla_l\nabla_i\phi
-\delta_{jl}\nabla_k\nabla_i\phi\pm
\epsilon_{ijkm}\nabla_l\nabla_m\phi
\mp\epsilon_{ijlm}\nabla_k\nabla_m\phi.}
It easily follows that
\eqn\axin{\grijkl=\mp {1\over 2} \epsilon_{kl}{}^{mn}
\hat R^i{}_{jmn}.}
So the (anti) self-duality appears in the gravitational sector
of the string
in terms of its generalized curvature thus justifying the name
``axionic
instanton" for the four-dimensional solution first found in
\refs{\rey,\reyone}.
For $e^{2\phi}=e^{2\phi_0}f$, where $f$ is given in \finst, we
obtain a
multi-instanton solution of string theory analogous to the YM
instanton.

In the special case of $e^{2\phi}=Q/r^2$, the sigma model
decomposes into the product of a one-dimensional CFT
and a three-dimensional WZW model with an $SU(2)$ group
manifold.
This can be seen by setting $u=\ln r$ and rewriting \sigmod\ in
 this case
in the form $I=I_1+I_3$, where
\eqn\onecft{I_1={1\over 4\pi\alpha'}\int d^2x
\left(Q(\partial u)^2
+\alpha' R^{(2)}\phi\right)}
is the action for a Feigin-Fuchs Coulomb gas, which is a
one-dimensional
CFT with central charge given by
$c_1=1+6\alpha'(\partial\phi)^2$ \gin. The imaginary charge of
 the
Feigin-Fuchs Coulomb gas describes the dilaton background growing
linearly in imaginary time.
$I_3$ is the Wess--Zumino--Witten \gepw\ action on an $SU(2)$
group manifold
with central charge
\eqn\threecharge{c_3={3k\over k+2}\simeq 3-{6\over k}+
{12\over k^2}+...}
where $k=Q/\alpha'$, called the ``level" of the WZW model, is an
integer. This can be seen from
the quantization condition on the Wess-Zumino term \gepw
\eqn\iwzw{\eqalign{I_{WZ}&={i\over 4\pi\alpha'}
\int_{\partial S^{\pm}_3}
d^2x\epsilon^{ab}\partial_ax^m\partial_bx^n B_{mn}\cr
&={i\over 12\pi\alpha'}\int_{S^{\pm}_3}d^3x\epsilon^{abc}
\partial_ax^m \partial_bx^n\partial_cx^p H_{mnp}\cr
&=2\pi i\left({Q\over\alpha'}\right).\cr}}
Thus $Q$ is not arbitrary, but is quantized in units of
$\alpha'$.
We use this splitting to obtain exact expressions
for the fields by fixing the metric and antisymmetric tensor
field
in their lowest order form and rescaling the dilaton order by
order in
$\alpha'$. The resulting expression for the dilaton is
\eqn\alldilaton{e^{2\phi}={Q\over r^{\sqrt{{4\over 1+
{2\alpha'\over Q}}}}}.}

The above bosonic solution easily generalizes to an analogous
solution
of heterotic string theory. The bosonic ansatz
\eqn\sansatz{\eqalign{
e^{-2\phi}\Box\ e^{2\phi}&=0,\cr
\met&=e^{2\phi}\delta_{mn}\qquad m,n=1,2,3,4\cr
g_{\mu\nu}&=\eta_{\mu\nu}\qquad\quad   \mu,\nu=0,5,6,7,8,9\cr
H_{mnp}&=\pm 2\epsilon_{mnpk}\partial^k\phi
\qquad m,n,p,k=1,2,3,4\cr}}
is a solution of the bosonic sector of the
ten-dimensional low-energy heterotic string effective action
\eqn\stringact{S_{10}={1\over 2\kappa_{10}^2} \int d^{10}x
\left(
R + 4(\partial\phi)^2 - {H^2\over 12}\right),}
whose equations of motion are equivalent to Weyl
invariance of the sigma-model. Eq.\sansatz\ with zero fermi
fields, zero gauge field and constant chiral spinor
$\epsilon=\epsilon_4 \otimes \eta_6$ in fact
preserves half the
spacetime supersymmetries stemming from the supersymmetry
equations
\eqn\sseq{\eqalign{\delta\psi_M&=\left(\partial_M+{\textstyle
{1\over 4}}\Omega_{MAB} \Gamma^{AB}\right)\epsilon=0,\cr
\delta\lambda&=\left(\Gamma^A\partial_A\phi-{\textstyle{1\over
12}} H_{ABC}\Gamma^{ABC}\right)\epsilon=0,\cr
\delta\chi&=F_{AB}\Gamma^{AB}\epsilon=0, \cr}}
where $A,B,C,M=0,1,2,...,9$, $\psi_M,\ \lambda$ and $\chi$ are
the gravitino, dilatino and gaugino fields and where
$\Omega_{M}{}^{PQ}=\omega_M{}^{PQ} - 1/2 H_M{}^{PQ}$ is the
generalized connection that generates the generalized curvature
\gcurv.
The (anti) self-duality of the generalized curvature \axin\ in
the
$(1234)$ subspace in fact
translates into an analogous condition for the generalized
connection
and is intimately connected to the choice of chirality of
$\epsilon_4$
that leads to
the preservation of precisely half of the supersymmetry
 generators.
In the above form \sansatz, we recover the tree-level
multi-fivebrane solution of \duflfb. The existence of the
fivebrane as a soliton solution of string theory lends support
to the string/fivebrane duality conjecture
\refs{\duf,\dufone,\str,\duflrsfd}, which states that the same
physics as superstring theory may be described by a theory
of fundamental superfivebranes propagating in ten dimensions.
This conjecture is a stringy analog of the Montonen-Olive
conjecture \mono, which postulates a duality between
electrically charged particles and magnetically charged
solitons in four-dimensional supersymmetric point field theory.

The simple expedient of
equating
the gauge connection $A_M{}^{PQ}$ to the generalized connection
 $\Omega_M{}^{PQ}$
then leads to another solution of \sseq, which possesses an
instanton in
both gauge and gravitational sectors. This solution was argued
to be an
exact solution of heterotic string theory (i.e. a solution to
all orders in $\alpha'$) which, in contrast to the purely
bosonic
solution above, does not require rescaling the dilaton
from its tree-level form \calhs. In fact, many of the pure
 gravity sector
solutions I will
discuss in this paper may be generalized to solutions with
 nontrivial
YM fields and which may be argued to be exact solutions of
heterotic
string theory by using the above gauge equals generalized
connection embedding
(first discovered in a somewhat different context in \chad).
For the sake
of simplicity, however, I will concentrate mainly on the former
class
of solutions and merely point out where such generalizations
 may be
interesting.

Another related axionic instanton
solution of heterotic string theory inspired by conformal field
theoretic
constructions in \koun\ is given by the
string-like solution \khubifb\
\eqn\bifbsol{\eqalign{e^{-2\phi_1}\Box\ e^{2\phi_1}&=
e^{-2\phi_2}\Box\ e^{2\phi_2}=0,\cr
\phi&=\phi_1 + \phi_2,\cr
\met&=e^{2\phi_1}\delta_{mn}\qquad m,n=2,3,4,5,\cr
g_{ij}&=e^{2\phi_2}\delta_{ij}\qquad i,j=6,7,8,9,\cr
g_{\mu\nu}&=\eta_{\mu\nu}\qquad\quad   \mu,\nu=0,1,\cr
H_{mnp}&=\pm 2\epsilon_{mnpq}\partial^q\phi
\qquad m,n,p,q=2,3,4,5,\cr
H_{ijk}&=\pm 2\epsilon_{ijkl}\partial^k\phi
\qquad i,j,k,l=6,7,8,9,\cr}}
which for constant chiral spinors
$\epsilon_\pm=\epsilon_2 \otimes \eta_4 \otimes \eta_4'$ solves
the
supersymmetry equations \sseq\ for zero fermi and gauge fields
(or alternatively for $A_M=\Omega_M$). In this case we have two
independent
axionic instantons, each of which breaks half the spacetime
supersymmetries. As a consequence, only $1/4$ of the original
supersymmetries are preserved. More recently, axionic instanton
solutions have been constructed in \bfrm.

\newsec{D=4 Solitons}


Let us single out a direction (say $x^4$) in the transverse
four-space $(1234)$
and assume all fields in \sansatz\ are independent of this
coordinate. Since
 all fields
are already independent of $x^5,x^6,x^7,x^8,x^9$, we may
 consistently assume
the $x^4,x^5,x^6,x^7,x^8,x^9$ are compactified on a
six-dimensional torus,
where we shall take the $x^4$ circle to have circumference
$Le^{-\phi_0}$
and the rest to have circumference $L$,
so that $\kappa_4^2=\kappa_{10}^2e^{\phi_0}/L^6$.
Going back to the 't Hooft ansatz \hfanstz, the solution
for $f$ satisfying the \ $f^{-1} \Box\ f=0$ has the form
\eqn\fmono{f_M=1+\sum_{i=1}^N{m_i\over |\vec x - \vec a_i|},}
where $m_i$ is proportional to the charge and $\vec a_i$ the
location in the
three-space $(123)$ of the $i$th instanton string. The solution
\sansatz\ with $e^{2\phi}=e^{2\phi_0} f_M$ can be reduced
to an explicit solution in the four-dimensional space $(0123)$
\dufkexst.
The reduction from ten to
five dimensions is trivial, as the metric is flat in the subspace
$(56789)$. In going from five to four dimensions, one follows the
 usual
Kaluza-Klein procedure \refs{\pol\grop{--}\sor}\ of replacing
 $g_{44}$ with a scalar field
$e^{-2\sigma}$. The tree-level effective action reduces in four
dimensions to
\eqn\redmon{I_4={1\over 2\kappa_4^2}\int d^4 x \sqrt{-g}
e^{-2\phi - \sigma}
\left( R + 4(\partial\phi)^2 + 4\partial\sigma\cdot\partial\phi -
e^{2\sigma} {M_{\alpha\beta}M^{\alpha\beta}\over 4} \right),}
where $\alpha,\beta=0,1,2,3$ and
where $M_{\alpha\beta}=H_{\alpha\beta 4}=\partial_\alpha
B_{\beta 4} -
\partial_\beta B_{\alpha 4}$. The four-dimensional monopole
solution for this
reduced action is then given by
\eqn\redmsol{\eqalign{e^{2\phi}=e^{-2\sigma}&=e^{2\phi_0}
\left(
1+\sum_{i=1}^N{m_i\over |\vec x - \vec a_i|}\right),\cr
ds^2&=-dt^2 + e^{2\phi}\left(dx_1^2 + dx_2^2 + dx_3^2\right),\cr
M_{ij}&=\pm\epsilon_{ijk}\partial_k e^{2\phi},\qquad i,j,k=1,2,3.\cr}}
For a single monopole, in particular, we have
\eqn\hmono{M_{\theta\phi}=\pm m \sin\theta,}
which is the magnetic field strength of a Dirac monopole. Note,
however, that
this monopole arose from
the compactified three-form $H$, and arises in all versions of
this
solution. In particular, one may obtain a multi-magnetic
monopole solution
of purely bosonic string theory \khuinst. A similar
reduction of instantons to monopoles was done in \back.

We now modify the solution of the 't Hooft ansatz even further
and choose
two directions in the four-space $(1234)$ (say $x^3$ and $x^4$)
and assume all
fields are independent of both of these coordinates. We may now
consistently
assume that $x^3,x^4,x^6,x^7,x^8,x^9$ are compactified on a
six-dimensional
torus, where we shall take the $x^3$ and $x^4$ circles to have
circumference
$Le^{-\phi_0}$ and the remainder to have circumference $L$, so
that
$\kappa_4^2=\kappa_{10}^2 e^{2\phi_0}/L^6$. Then the solution
for $f$ satisfying $f^{-1} \Box\ f=0$ has multi-string
structure
\eqn\fstring{f_S=1-\sum_{i=1}^N \lambda_i \ln
|\vec x - \vec a_i|,}
where $\lambda_i$ is the charge per unit length and $\vec a_i$
the location in
the two-space $(12)$ of the $i$th string.
By setting $e^{2\phi} = e^{2\phi_0}f_S$, we obtain from
\sansatz\ a
multi-string solution which reduces to a solution in the
four-dimensional
space $(0125)$. The reduction from
ten to six dimensions is trivial, as the metric is flat in the
subspace
$(6789)$. In going from six to four dimensions, we compactify
the $x_3$ and
$x_4$
directions and again follow the Kaluza-Klein procedure by
 replacing $g_{33}$
and $g_{44}$ with a scalar field $e^{-2\sigma}$. The tree-level
effective
action reduces in four dimensions to
\eqn\redst{S_4={1\over 2\kappa_4^2}\int d^4 x \sqrt{-g}
 e^{-2\phi - 2\sigma}
\left( R + 4(\partial\phi)^2 + 8\partial\sigma\cdot\partial\phi
+
2(\partial\sigma)^2 - e^{4\sigma} {N_\rho N^\rho\over 2}
\right),}
where $\rho=0,1,2,5$, where $N_\rho=H_{\rho 34}=
\partial_\rho B$, and where
$B=B_{34}$. The four-dimensional
string soliton solution for this reduced action is then given
by
\dufkexst
\eqn\redssol{\eqalign{e^{2\phi}=e^{-2\sigma}&=e^{2\phi_0}\left(
1-\sum_{i=1}^N \lambda_i \ln |\vec x - \vec a_i|\right),\cr
ds^2&=-dt^2 + dx_5^2 + e^{2\phi}\left(dx_1^2 + dx_2^2\right),
\cr
N_i&=\pm \epsilon_{ij} \partial_j e^{2\phi}.\cr}}

We complete the family of solitons that can be obtained from
the
solutions
of the 't Hooft ansatz by demanding that $f$ depend on only one
coordinate,
say $x^1$. We may now consistently assume that
$x^2,x^3,x^4,x^7,x^8,x^9$ are
compactified on a six-dimensional torus, where we shall take
the $x^2$,
$x^3$ and $x^4$ circles to have circumference $Le^{-\phi_0}$
and the rest to
have circumference $L$, so that
$\kappa_4^2=\kappa_{10}^2e^{3\phi_0}/L^6$.
Then the solution
of $f^{-1} \Box\ f=0$ has domain wall structure with the
``confining potential"
\eqn\fdom{f_D=1+ \Lambda x_1,}
where $\Lambda$ is a constant. By setting
$e^{2\phi} = e^{2\phi_0} f_D$ in \sansatz, we obtain a
multi-domain wall
solution which once more can be explicitly reduced to
four dimensions.
For the spacetime $(0156)$, the tree-level effective
action in $D=4$ has the form
\eqn\reddom{S_4={1\over 2\kappa^2}\int d^4 x \sqrt{-g}
e^{-2\phi - 3\sigma}
\left( R + 4(\partial\phi)^2 + 12\partial\sigma\cdot\partial\phi +
6(\partial\sigma)^2 - e^{6\sigma} {P^2\over 2} \right),}
where $P$ is a cosmological constant. Note that \reddom\ is not
 obtained by
a simple reduction of the ten-dimensional action owing to the
nonvanishing
of $H_{234}$. The four-dimensional domain wall
solution for this reduced action is then given by \dufkexst
\eqn\reddomsol{\eqalign{e^{2\phi}=e^{-2\sigma}&=
e^{2\phi_0}\left(1+\sum_{i=1}
^N\Lambda_i |x_1-a_i|\right),\cr
ds^2&=-dt^2 + dx_5^2 + dx_6^2 + e^{2\phi} dx_1^2,\cr
P&=\sum_{i=1}^N\Lambda_i \left(\Theta(x_1-a_i)-\Theta(-x_1+a_i)
\right).\cr}}
A trivial change of coordinates reveals that the spacetime is,
in fact, flat.
Dilatonic domain walls with a flat spacetime have been discussed
in a somewhat
different context in \cvet.

For both strings and domain walls, generalizations to solutions
with
$A_M=\Omega_M$ are straightforward \refs{\khumonex,\dufkexst}.
In all three cases, the
multi-soliton solutions break half the spacetime supersymmetries of
the $N=4$ four-dimensional heterotic string theory to which the
original $N=1$ ten-dimensional heterotic string theory is
toroidally
compactified. The existence of these solutions owes to the
 saturation
of a Bogomol'nyi bound between mass per unit volume and
topological
charge, and which results in a ``zero-force'' condition
analogous to
that found for BPS monopoles \refs{\pras,\bog}.

\newsec{String/String Duality}

Let us focus on the solitonic string configuration \redssol\ in
the case of
a single source \dufkexst. In terms of the complex field
\eqn\tfield{\eqalign{T&=T_1+iT_2\cr &=e^{-2\sigma}-iB_{34} \cr
&=\sqrt{{\rm det} g^S_{mn}}-iB_{34}
\qquad m,n=3,4,\cr}}
where $g^S_{MN}$ is the string $\sigma$-model metric, the
 solution takes the
form (with $z=x_1+x_2$)
\eqn\tsol{\eqalign{T&=-{1\over 2\pi}\ln {z\over r_0},\cr
ds^2&=-dt^2 + dx_5^2 - {1\over 2\pi} \ln{r\over r_0} dz
d\overline z,\cr}}
whereas both the four-dimensional (shifted) dilaton $\eta=\phi
 + \sigma$
and the four-dimensional two-form $B_{\mu\nu}$ are zero. In
terms of the
canonical metric $g_{\mu\nu}$, $T_1$ and $T_2$, the relevant
part of the
action takes the form
\eqn\sgtt{S_4={1\over 2\kappa_4^2}\int d^4 x\sqrt{-g}
\left( R - {1\over 2T_1^2}g^{\mu\nu}\partial_\mu T
\partial_\nu \overline T
\right)}
and is invariant under the $SL(2,R)$ transformation
\eqn\sltwor{T\to {aT+b\over cT+d},\qquad ad-bc=1.}
The discrete subgroup $SL(2,Z)$, for which $a$, $b$, $c$ and
$d$ are
integers, is just a subgroup of the $O(6,22;Z)$ {\it target
space duality},
which can be shown to be an exact symmetry of the compactified
string theory
at each order of the string loop perturbation expansion
\refs{\kikyam\sakai\bush\nair\duff\tseyv\tseycqg{--}\givpr}.

This $SL(2,Z)$ is to be contrasted with the $SL(2,Z)$ symmetry
of the
elementary four-dimensional solution of Dabholkar {\it et al.}
\dabghr.
In the latter solution $T_1$ and $T_2$ are zero, but $\eta$ and $B_{\mu\nu}$
are non-zero. The relevant part of the action is
\eqn\dabactt{S_4={1\over 2\kappa_4^2}\int d^4 x\sqrt{-g}
\left( R - 2g^{\mu\nu}\partial_\mu \eta \partial_\nu \eta
-{1\over 12} e^{-4\eta} H_{\mu\nu\rho} H^{\mu\nu\rho}
\right).}
The equations of motion of this theory also display an
$SL(2,R)$ symmetry,
but this becomes manifest only after dualizing and introducing
the axion
field $a$ via
\eqn\axfield{\sqrt{-g}g^{\mu\nu}\partial_\nu a=
{1\over 3!}\epsilon^{\mu\nu\rho\sigma} H_{\nu\rho\sigma}
e^{-4\eta}.}
Then in terms of the complex field
\eqn\comps{\eqalign{S&=S_1 + iS_2 \cr
&=e^{-2\eta} + ia \cr}}
the Dabholkar {\it et al.} fundamental string solution may be
written
\eqn\dabs{\eqalign{S&=-{1\over 2\pi} \ln {z\over r_0},\cr
ds^2&=-dt^2 + dx_5^2 - {1\over 2\pi} \ln {r\over r_0} dz
d\overline z.\cr}}
Thus \tsol\ and \dabs\ are the same with the replacement
$T\leftrightarrow
S$. It has been conjectured that this second $SL(2,Z)$ symmetry
may also be a
symmetry of string theory \refs{\fonilq,\senone,\schtwo,\vafw},
but
this is far from
obvious order by order in the string loop expansion since it
involves a
strong/weak coupling duality $\eta\to - \eta$. What
interpretation
are we to give to these two $SL(2,Z)$ symmetries: one an
obvious symmetry of
the fundamental string and the other an obscure symmetry of
the
fundamental
string?

Related issues are brought up
by Sen \sentwo, Schwarz and Sen \schsen\
and
Bin\'etruy
\bin. In particular, Sen draws attention to the Dabholkar
{\it et al.} string
solution \dabs\ and its associated $SL(2,Z)$ symmetry as
supporting evidence
in favor of the conjecture that $SL(2,Z)$ invariance may indeed
be an exact
symmetry of string theory. He also notes
that the spectrum of electric and magnetic charges is
consistent with the
proposed $SL(2,Z)$ symmetry \sentwo.\footnote{$^\dagger$}
{Sen also discusses the concept of a ``dual string", but for
him this is
obtained from the fundamental string by an $SL(2,Z)$ transform.
For us, a
dual string is obtained by the replacement
$S\leftrightarrow T$.}

All of these observations fall into place if one accepts the
proposal of
Schwarz and Sen \schsen: {\it under string/fivebrane duality
the roles of
the target-space duality and the strong/weak coupling duality
are
interchanged !} This proposal is entirely consistent with an
earlier one that
under string/fivebrane duality the roles of the $\sigma$-model
loop
expansion and the string loop expansion are interchanged
\duflloop. In this
light, the two $SL(2,Z)$ symmetries discussed above are just
what one
expects. From the string point of view, the $T$-field $SL(2,Z)$
is an
obvious target space symmetry, manifest order by order in
string loops
whereas the $S$-field $SL(2,Z)$ is an obscure strong/weak
coupling symmetry.
{}From the fivebrane point of view, it is the $T$-field $SL(2,Z)$
which is
obscure while the $S$-field $SL(2,Z)$ is an ``obvious" target
space
symmetry.
(This has not yet been proved except at the level of the
low-energy field
theory, however. It would be interesting to have a proof
starting from the
worldvolume of the fivebrane.)
This interchange in the roles of the $S$ and $T$ field in
going from the string to the fivebrane has also been noted by
Bin\'etruy
\bin. It is made more explicit when $S$ is expressed in terms
of the
variables appearing naturally in the fivebrane version
\eqn\fbvers{\eqalign{S&=S_1 + iS_2 \cr
&=e^{-2\eta}+ia_{034789},\cr
&=\sqrt{{\rm det} g^F_{mn}}+ia_{034789},
\qquad m,n=3,4,6,7,8,9, \cr}}
where $g^F_{MN}=e^{-\phi/3}g^S_{MN}$ is the fivebrane
$\sigma$-model metric
\duflrsfd\ and
$a_{MNPQRS}$ is the 6-form which couples to the 6-dimensional
worldvolume of
the fivebrane, in complete analogy with \tfield.

It may at first sight seem strange that a string can be dual to
another
string in $D=4$. After all, the usual formula relating the
dimension of an
extended object, $d$, to that of the dual object, $\tilde d$,
is
$\tilde d=D-d-2$. So one might expect string/string duality
only in $D=6$
\duflloop. However, when we compactify $n$ dimensions and allow
the dual object to wrap around $m\leq d-1$ of the compactified
directions
we find $\tilde d_{{\rm effective}}=\tilde d -m=D_{{\rm
effective}}-d-2
+(n-m)$, where $D_{{\rm effective}}=D-n$. In particular for
$D_{{\rm
effective}}=4$, $d=2$, $n=6$ and $m=4$, we find $\tilde d_{{\rm
effective}}=2$.

Thus the whole string/fivebrane duality conjecture is put in a
different
light when viewed from four dimensions. After all, our
understanding of the
quantum theory of fivebranes in $D=10$ is rather poor, whereas
the quantum
theory of strings in $D=4$ is comparatively well-understood
(although we
still have to worry about the monopoles and domain walls). In
particular,
the dual string will presumably exhibit the normal kind of mass
spectrum
with linearly rising Regge trajectories, since the classical
($\hbar$-independent) string expression $\widetilde T_6 L^4
\times
({\rm angular\ momentum})$ has dimensions of $({\rm mass})^2$,
whereas
the analogous classical expression for an uncompactified
fivebrane is
$(\widetilde T_6)^{1/5} \times ({\rm angular\ momentum})$ which
has
dimensions $({\rm mass})^{6/5}$ \duf. Indeed, together with the
observation
that the $SL(2,Z)$ strong/weak coupling duality appears only
after
compactifying at least $6$ dimensions, it is tempting to revive
the earlier
conjecture \refs{\duf,\fujku}\ that the internal consistency of
 the
fivebrane may actually {\it require} compactification.
For recent discussions of the string/string duality
conjecture see \refs{\dufm\glasgow{--}\dufnew}).

Since the full $T$-duality corresponds to the discrete group
$O(6,22;Z)$
and is known to be an
exact symmetry of the full string theory,
the dual string of \dufkexst\
thus belongs to an $O(6,22;Z)$ family of dual
strings just as
there is an $SL(2,Z)$ family of fundamental strings \sentwo.
In recent work, Sen \sentd\ discussed the issue of whether
$S$ duality and $T$ duality can be combined
in a larger duality group (see also \dufr).
In particular, he argued that in heterotic string theory
compactified on a seven-torus the $SL(2,Z)$ $S$ duality
can be combined with the $O(7,23;Z)$ $T$ duality into the group
$O(8,24;Z)$. The existence of a Killing direction in all of the
solitons discussed in this paper means that they may all be
viewed as point-like solutions of three-dimensional heterotic
string theory. Thus from the viewpoint of three-dimensional
heterotic string theory $O(8,24;Z)$ appears as a duality group,
whereas from the point of view of four-dimensional heterotic
string
theory it appears as a solution-generating group.

Central to the issue of combining $S$ and $T$ duality is whether
there exists an $O(8,24;Z)$  transformation that
maps the fundamental $S$ string of \dabghr\ to the dual
$T$ string of \dufkexst.
In \sentd, a transformation is found that takes the
fundamental string to the stringy ``cosmic'' string solution
of \gresvy. This latter solution, however, while very
interesting
(and nonsingular at large distances
in constrast to both the $S$ string and the $T$
string), does not, unlike the dual string,
represent the fundamental solution of
a (conjectured) dual theory. The explicit $O(8,24;Z)$
transformation that maps the $S$ string into the $T$ string is
given as follows.
Following Sen's notation (see \sentd), ${\cal M}_T$, the
$32\times 32$ matrix that corresponds to the $T$ string, is
obtained from ${\cal M}_S$, the
$32\times 32$ matrix that corresponds to the $S$ string,
simply by
exchanging (rows and columns) $1$ with $10$, $2$ with $31$,
$3$ with $8$ and $9$ with $32$. The transformation matrix is
therefore, effectively, an  $O(4,4;Z)$ matrix.
It follows that
all three strings, fundamental, dual and cosmic are
related by $O(8,24;Z)$, but only the first two are relevant
for the purposes of combining $S$ and $T$ duality
(for more details see \duffkr). An interesting discussion
of the connections between duality groups can be found in \hult.

Repeating Sen's and other arguments on three-dimensional
reduction for $N=2$ superstring theory, we can infer that
the larger duality group (or solution-generating
group) for $N=2$ containing $S$ and $T$ duality
is connected to a dual quaternionic manifold. In the case
of $n$ moduli, this group is $SO(n+2,4;Z)$
\refs{\cecfg,\ferquat}.

\newsec{D=4 Black Holes}

 We now extend the three solutions of section 3 to two-parameter
solutions of the low-energy
equations of the four-dimensional heterotic string \dufkmr,
characterized by a mass
per unit $p$-volume, ${\cal M}_{p+1}$, and magnetic charge,
$g_{p+1}$, where $p = 0, 1$ or $2$.
The solitons
discussed in section 3 are recovered
in the extremal limit,
 $ \sqrt{2} \kappa {\cal M}_{p+1} =  g_{p+1}$. The
two-parameter solution extending the supersymmetric monopole
corresponds to
a magnetically charged black hole, while the solution extending
 the
supersymmetric domain wall corresponds to a black membrane. By
 contrast, the
two-parameter string solution does not possess a finite horizon
 and
corresponds to a naked singularity.

All three solutions involve both the
dilaton and the modulus fields, and are thus to be contrasted
with pure dilaton
solutions. In particular, the effective scalar coupling to the
Maxwell field,
$e^{-\alpha\phi} F_{\mu\nu} F^{\mu\nu}$, gives rise to a new
string black hole
with $\alpha = \sqrt{3}$, in contrast to the pure dilaton
solution of the
heterotic string which has $\alpha = 1$
\refs{\gib\gibm\ghs\shatw\klopv\kal\ko\hor\senprl{--}\gidps}.
It thus resembles the black hole
previously studied in the context of Kaluza-Klein theories
\refs{\dobm,\chod,\pol\grop{--}\sor,\gibp,\gib,\hor} which also
has $\alpha = \sqrt{3}$, and which reduces to the
Pollard-Gross-Perry-Sorkin
\refs{\pol\grop{--}\sor} magnetic monopole in the extremal limit.
In this connection, we recall \hw, according to which
$\alpha > 1$ black holes behave like elementary particles!

The fact that the heterotic string admits
$\alpha = \sqrt{3}$ black holes also
has implications for string/fivebrane duality
\refs{\duf\str{--}\duflrsfd}.
We shall show that electric/magnetic duality in $D=4$ may be
seen as a
consequence of string/fivebrane
duality in $D=10$.

We begin with the two-parameter black hole.
The solution of the action \redmon\ is given by
\eqn\bmonrs{\eqalign{e^{-2\Phi}&=e^{2\sigma_{1}}=
\left(1 - {r_-\over r}
\right),\cr
ds^2&=-\left(1-{r_+\over r}\right)
\left(1-{r_-\over r}\right)^{-1}dt^2 +
\left(1-{r_+\over r}\right)^{-1}dr^2 +
r^2\left(1-{r_-\over r}\right)d\Omega_2^2,\cr
F_{\theta\varphi}&=\sqrt{r_+r_-} \sin\theta \cr}}
where here, and throughout this section, we set the dilaton vev
 $\Phi_0$ equal
to zero. This represents a magnetically charged black hole with
event horizon at $r=r_+$ and inner horizon at $r=r_-$.
The magnetic charge and mass of the black hole are given by
\eqn\massch{\eqalign{g_{1}&={4\pi\over {\sqrt{2}\kappa}}
(r_+r_-)^{{1}\over{2}},
   \cr
{\cal M}_1&={{2 \pi}\over {{\kappa}^2}}(2r_+ - r_-) .\cr}}
Changing coordinates via $y=r-r_-$ and taking the extremal
limit $r_+=r_-$
yields:
\eqn\monex{\eqalign{e^{2\Phi}&=e^{-2\sigma_{1}}=\left(1 +
{r_-\over y}
\right),\cr
ds^2&=-dt^2 + e^{2\Phi}\left(dy^2 + y^2d\Omega_2^2\right),\cr
F_{\theta\varphi}&=r_- \sin\theta.\cr}}
which is just the supersymmetric
monopole solution of section 3 which saturates
the Bogomol'nyi bound $\sqrt{2} \kappa{\cal M}_1\ge g_1$.

Next we derive a two-parameter string solution which, however,
does not possess a finite event horizon and consequently cannot
 be
interpreted as a black string. A two-parameter family of
solutions of the
action \redst\ is now
given by
\eqn\stsol{\eqalign{e^{2\Phi}&=e^{-2\sigma_{2}}=
(1+k/2-\lambda \ln y),\cr
ds^2&=-(1+k)dt^2 + (1+k)^{-1}(1+k/2-\lambda \ln y)dy^2 +
y^2(1+k/2-\lambda \ln y)d\theta^2 + dx_3^2,\cr
F_\theta&=\lambda \sqrt{1+k},\cr}}
where for $k=0$ we recover the supersymmetric string soliton
solution of
section 3 which, as shown in section 4, is dual to the
elementary string
solution of Dabholkar {\it et al.}. The solution shown in
\stsol\ in fact
represents a
naked singularity,
since the event horizon is pushed out to $r_+=\infty$, which
agrees with
the Horowitz-Strominger ``no-$4D$-black-string'' theorem \hors.

Finally, we consider the two-parameter black membrane solution.
The two-parameter black membrane solution of the action
\reddom\ is then
\eqn\bmemrs{\eqalign{e^{-2\Phi}&=e^{2\sigma_{3}}=
\left(1-{r\over r_-}
\right),\cr
ds^2&=-\left(1-{r\over r_+}\right)\left(1-{r\over r_-}\right)^
{-1}dt^2 +
\left(1-{r\over r_+}\right)^{-1}\left(1-{r\over r_-}\right)^
{-4}dr^2 +
dx_2^2 + dx_3^2,\cr
F&=-(r_+r_-)^{-1/2}.\cr}}
This solution represents a black membrane with event horizon at
$r=r_+$ and inner horizon at $r=r_-$.
Changing coordinates via $y^{-1}=r^{-1}-r_{-}^{-1}$ and taking
the extremal
limit yields
\eqn\domex{\eqalign{e^{2\Phi}&=e^{-2\sigma_3}=\left(1+
{y\over r_-}
\right),\cr
ds^2&=-dt^2 + dx_2^2 + dx_3^2 + e^{2\Phi} dy^2,\cr
F&=-{1\over r_-}.\cr}}
which is just the supersymmetric domain wall solution of
 section 3.

 We note that the black hole solution corresponds to a
Maxwell-scalar coupling $e^{-\a \phi} F_{\mu \nu} F^{\mu \nu}$
with
$\a=\sqrt{3}$. This is to be contrasted with
the pure dilaton black hole solutions of the heterotic string
that have attracted much attention
recently
\refs{\gib\gibm\ghs\shatw\klopv\kal\ko\hor\senprl{--}\gidps}
and have $\alpha=1$
\footnote{$^\ast$}{Contrary to some claims in the
literature, the pure Reissner-Nordstr\"om
black hole with $\a=0$ is also a solution of the low energy
heterotic string
equations. This may be seen by noting that it provides a
solution to $(N=2,
\, D=4)$ supergravity which is a consistent truncation of
 toroidally
compactified $N=1, \, D=10$ supergravity \dupo.}. The case
$\alpha=\sqrt{3}$
also occurs when
the Maxwell field and the scalar field $\phi$ arise
from a Kaluza-Klein reduction of
pure gravity from $D=5$ to $D=4$:
\eqn\redmet {\hat{g}_{MN} = e^{\phi\over{\sqrt{3}}}
   \pmatrix { g_{\mu\nu}+e^{-\sqrt{3} \phi} A_{\mu}A_{\nu} &
                    e^{-\sqrt{3} \phi} A_{\mu} \cr
          e^{-\sqrt{3} \phi} A_{\nu} &   e^{-\sqrt{3} \phi} ,
\cr}}
where $\hat{g}_{MN} \, (M,N=0,1,2,3,4)$ and $g_{\mu\nu} \,
(\mu,\nu=0,1,2,3)$ are the canonical metrics in 5 and 4
 dimensions
respectively. The resulting action is given by
\eqn\redact{S={1\over2 \kappa^2} \int d^4 x \sqrt{-g}
 \left[ R -{1\over 2}
   (\partial \phi)^2 -{1\over 4} e^{-\sqrt{3} \phi} F_{\mu \nu}
F^{\mu \nu}
   \right],}
and it admits as an ``elementary'' solution the $\a=\sqrt{3}$
black hole
metric \bmonrs,  but with the scalar field
\eqn\die{e^{-2 \phi}= \Delta_{-}^{\sqrt{3}}}
and the electric field
\eqn\fse{{1\over \sqrt{2} \kappa}e^{-\sqrt{3} \phi \, \ast}
F_{\theta\varphi}=
         {e\over 4 \pi} \sin\theta}
corresponding to an electric monopole with Noether charge $e$.
This system also admits the topological magnetic solution with
\eqn\dim{e^{-2 \phi}= \Delta_{-}^{-\sqrt{3}}}
and the magnetic field
\eqn\fsm{{{1}\over{\sqrt{2}\kappa}} F_{\theta\varphi}=
{g\over4 \pi}\sin\theta}
corresponding to a magnetic monopole with  topological magnetic
charge $g$ obeying the Dirac quantization rule
\eqn\diqua{eg=2 \pi n, \, \, \, n={\rm integer.}}
In effect, it was for this reason that the $\alpha=\sqrt{3}$
black hole
was identified as a solution of the Type II string in
\duflblacks, the fields
$A_\mu$ and $\phi$ being just the abelian gauge field and the
dilaton of
($N=2,\, D=10$) supergravity which arises from Kaluza-Klein
compactification
of ($N=1,\, D=11$) supergravity.

Some time ago, it was pointed out in \gibp\ that $N=8$
supergravity,
compactified from $D=5$ to $D=4$, admits an infinite tower of
elementary states
with mass $m_n$ and electric charge $e_n$ given by $\sqrt{2}
\kappa m_n=e_n$,
where $e_n$ are quantized in terms of a fundamental charge $e$,
 $e_n=n \, e$,
and that these elementary states fall into $N=8$ supermultiplets.
 They also
pointed out that this theory admits an infinite tower of
solitonic states
with the masses  $\tilde{m}_n$ and magnetic  charge $g_n$ given
by
$\sqrt{2}\kappa\tilde{m}_n=g_n=n\, g$, where $e$ and $g$ obey
$eg=2 \pi$,
which also fall into the same $N=8$ supermultiplets. The
authors of \gibp\
 conjectured,
$\acute{\rm a}$ la Olive-Montonen \mono, that there should
exist a dual formulation of the theory for which the roles of
electric
elementary states and magnetic solitonic states are
interchanged.
It was argued in \duflblacks\ that this electric/magnetic
duality conjecture
in $D=4$ could be reinterpreted as a particle/sixbrane duality
conjecture
in $D=10$.

To see this, consider the action dual to $S$, with
$\a=-\sqrt{3}$, for
which the roles of Maxwell field equations and Bianchi
identities are
interchanged:
\eqn\duact{\tilde{S}=
        {1\over2 \kappa^2} \int d^4 x \sqrt{-g} \left[ R -
{1\over 2}
   (\partial \phi)^2 -{1\over 4} e^{\sqrt{3} \phi}
      \tilde{F}_{\mu \nu} \tilde{F}^{\mu \nu}
   \right],}
where $\tilde{F}_{\mu\nu} =e^{-\sqrt{3} \phi \, \ast}
F_{\mu\nu} .$
This is precisely the action obtained by double dimensional
reduction of
a dual formulation of ($D=10, \, N=2$) supergravity in which
the two-form
$F_{MN}$ ($M,N=0,...,9$) is swapped for an 8-form
$\tilde{F}_{M_1..M_8}$, where $\tilde{F}_{\mu\nu}=
\tilde{F}_{\mu\nu456789}$.
This dual action also admits both electric and magnetic
monopole solutions
but because the roles of field equations and Bianchi identities
are
interchanged, so are the roles of electric and magnetic. Since
the 1-form
and 7-form potentials, which give rise to these 2-form and
8-form
field strengths, are those that couple naturally to the
worldline of
a point particle or the worldvolume of a 6-brane, we see that
the
Gibbons-Perry ($N=8,\, D=4$) electric/magnetic duality
conjecture may be
re-expressed as an (Type II, $D=10$) particle/sixbrane duality
conjecture.
Indeed, the $D=10$ black sixbrane of \hors\ is simply obtained
by adding 6 flat dimensions to the $D=4,\, \alpha=\sqrt{3}$
magnetic black hole.

In general, the $N=8$ theory will admit black holes with
$\a=0,1$ and
$\sqrt{3}$ whose extreme limits preserve $1,2$ or $4$ spacetime
supersymmetries,
respectively. Defining ${\cal M}_1=M,\ g_1^2=4\pi Q^2$ and
$\kappa^2
=8\pi G$, these extreme black holes satisfy the ``no-force''
condition,
i.e. they saturate the Bogomol'nyi bounds
\eqn\bogo{G(M^2+\Sigma^2)=(1+\a^2)GM^2=N'GM^2=Q^2}
where $\Sigma=\a M$ is the scalar charge and $N'$ is the number
of unbroken
supersymmetries.

The solutions presented in this section now allow us to discuss
the
$\a=\sqrt{3}$
electric/magnetic duality from a totally different perspective
from above.
For concreteness, let us focus on  generic toroidal
compactification of the
heterotic string. Instead of $N=8$ supergravity, the
four-dimensional theory is now
$N=4$ supergravity coupled to 22  $N=4$ vector
multiplets\footnote{$^\dagger$}
{Gibbons discusses both the $\alpha=1$ black hole of pure $N=4$
supergravity and the $\alpha=\sqrt{3}$ Kaluza-Klein black hole
in the same paper
\gib, as does Horowitz \hor.
Moreover, black
holes in pure $N=4$ supergravity are treated by Kallosh {\it
et
al.}
\refs{\klopv\kal{--}\ko}.
The reader may therefore wonder why the $\a=\sqrt{3}$ $N=4$
black hole
discussed in the present paper was overlooked. The reason is
that pure $N=4$
supergravity does not admit the $\a=\sqrt{3}$ solution; it is
crucial that
we include the $N=4$ vector multiplets in order to introduce
the modulus
fields.}.
The same dual Lagrangians \redact\ and \duact\ still emerge but
with
completely different origins. The Maxwell field $F_{\mu\nu}$
(or $\tilde{F}_
{\mu\nu}$) and the scalar field $\phi$ do not come from the
$D=10$ $2$-form (or
$8$-form) and dilaton of the Type II particle (or sixbrane),
but rather from
the $D=10$ $3$-form (or $7$-form) and dilaton plus modulus
field of the
heterotic string (or heterotic fivebrane). Thus, the $D=4$
electric/magnetic
duality can now be re-interpreted as a $D=10$ string/fivebrane
duality!

Because of the non-vanishing modulus field
$g_{44}=e^{-2\sigma}$ however,
the $D=10$ black fivebrane solution is not obtained by adding 6
flat
dimensions to the $D=4$ black hole. Rather the two are
connected by wrapping
the fivebrane around 5 of the 6 extra dimensions \dufkexst.

Of course, we have established only that these two-parameter
configurations are solutions of the field theory limit of the
heterotic string.
Although the extreme one-parameter solutions are expected to be
exact to all
orders in $\alpha'$, the same reasoning does not carry over to
the new
two-parameter solutions.

It would be also interesting to see whether the generalization
of the
one-parameter solutions of section 3 to the two-parameter
solutions of this
section can be carried out when we include the Yang-Mills
coupling.
 This would necessarily involve giving up
the self-duality condition on the Yang-Mills field strength,
however, since the
self-duality condition is tied to the extreme, $\sqrt{2} \kappa
{\cal M}_{p+1}
=  g_{p+1}$, supersymmetric solutions.

\newsec{Dynamics of String Solitons}

All the soliton solutions we have discussed so far have the
property, like BPS
magnetic monopoles, that they exert zero static force on
each other and can be superposed to form multi-soliton
solutions with
arbitrarily variable collective coordinates. Since these static
properties are also possessed by fundamental strings winding
around an
infinitely large compactified dimension, it was conjectured in
\dabghr\ that the
elementary string is actually the exterior solution for an
infinitely long
fundamental string. In this section we show that, in
contradistinction to the BPS case, the velocity-dependent forces
between these string solitons also vanish (i.e. we argue that
the
scattering is trivial).  We also argue that this phenomenon
provides further,
dynamical evidence for the identification of the elementary
string solution
with the underlying fundamental string by comparing the
scattering of the
elementary
solutions with expectations from a Veneziano amplitude
computation for
macroscopic fundamental strings \khuscat.

As shown in \calk, the static ansatz leads to a vanishing
leading-order velocity-dependent force to for a test string
propagating in the
background
of an elementary string. This result also holds for the other
soliton solutions we have discussed in this paper.
In particular,
test monopoles propagating in the background of a source monopole
also do not experience a dynamic force to leading order.
As this is a rather surprising result, we would like to compute
the metric on
moduli space for these solitons.
  The geodesics of this metric represent the motion of
quasi-static solutions
in the static solution manifold and in the absence of a full
time dependent
solution provide a good approximation to the low-energy
dynamics of the
solitons. In all cases the metric is found to be flat in
agreement with the
test-soliton approximation, which again implies vanishing
dynamical force in
the low-velocity limit. Here we summarize the computation for
the metric on
moduli space for monopoles discussed in
\refs{\khumonscat,\khumonex}.

Manton's prescription \mantwo\ for the study of soliton
scattering may be
summarized as follows. We first invert the constraint equations
of the system.
The resultant time dependent field configuration does not in
general
satisfy the full time dependent field equations, but provides
an initial data point for the fields and their time derivatives.
Another way of saying this is that the initial motion is
tangent to the
set of exact static solutions.  The kinetic action obtained
by replacing the solution to the constraints into the action
defines a
metric on the parameter space of static solutions. This metric
defines
geodesic motion on the moduli space \mantwo.

A calculation of the metric on moduli space for the scattering
of BPS
monopoles and a description of its geodesics was worked out by
Atiyah
and Hitchin \atihone. Several interesting properties of monopole
scattering were found, such as the conversion of monopoles into
dyons
and the right angle scattering of two monopoles on a direct
collision
course \refs{\atihone,\atihtwo}. The configuration space is
found to
be a four-dimensional manifold $M_2$ with a self-dual Einstein
metric.

Here we adapt Manton's prescription to study the dynamics of
the heterotic
 string monopoles discussed in section 3. We follow essentially
the same steps that Manton outlined for monopole scattering,
but take
into account the peculiar nature of the string effective
action. Since
we work in the low-velocity limit, our kinematic analysis is
nonrelativistic.

We first solve the constraint equations for the monopoles.
These equations are simply the $(0j)$
components of the tree-level equations of motion
\eqn\constraints{\eqalign{R_{0j}-{1\over 4}H^2_{0j}+
2\nabla_0\nabla_j\phi&=0,
\cr
-{1\over 2}\nabla_kH^k{}_{0j}+H_{0j}{}^k\partial_k\phi&=0\cr}}
which follow from the action \stringact.
We wish to find an $O(\beta)$ solution to the above equations
which
represents a quasi-static version of the neutral multi-monopole
solution (i.e.
 a multi-monopole solution with time dependent $\vec a_i$).
Here we use the uncompactified solution \sansatz\ with
$e^{2\phi}=e^{2\phi_0}f_M$, with $f_M$ given in \fmono, as
opposed to the
explicitly compactified version \redmsol\ (in particular, we do
not make the
replacement $g_{44}=e^{-2\sigma}$) although the results in both
cases are
identical, in order to more easily keep track of the terms in
the former case.
We give each monopole an arbitrary transverse velocity
$\vec\beta_n$ in the $(123)$ subspace of the four-dimensional
transverse space
and see what corrections to the fields are required by the
constraints. The vector $\vec a_n$ representing the position of
the
$n$th monopole in the three-space $(123)$ is given by
\eqn\aunty{\vec a_n(t)=\vec A_n + \vec\beta_nt,}
where $\vec A_n$ is the initial position of the $n$th monopole.
Note that at
$t=0$ we recover the exact static multi-monopole solution. Our
solution to
the constraints will adjust our quasi-static approximation so
that the
initial motion in the parameter space is tangent to the initial
exact solution at $t=0$.
The $O(\beta)$ solution to the constraints is given by
\khumonscat
\eqn\orderbeta{\eqalign{e^{2\phi(\vec x,t)}&=1+\sum_{n=1}^N{m_n
\over
|\vec x - \vec a_n(t)|},\cr g_{00}&=-1,\qquad g^{00}=-1,
\qquad
g_{ij}=e^{2\phi}\delta_{ij},\qquad g^{ij}=e^{-2\phi}\delta_{ij},
\cr
g_{0i}&=-\sum_{n=1}^N{m_n\vec\beta_n\cdot \hat x_i\over
|\vec x - \vec
a_n(t)|},\qquad g^{0i}=e^{-2\phi}g_{0i},\cr
H_{ijk}&=\epsilon_{ijkm}\partial_m e^{2\phi},\cr
H_{0ij}&=\epsilon_{ijkm}\partial_m g_{0k}=\epsilon_{ijkm}
\partial_k
\sum_{n=1}^N{m_n\vec\beta_n\cdot \hat x_m\over |\vec x -
\vec a_n(t)|},\cr}}
where $i,j,k,m=1,2,3,4$, all other metric components are flat,
all other
components of $H$ vanish, the $\vec a_n(t)$ are given by
\aunty\ and we use a
flat space $\epsilon$-tensor. Note that $g_{00}$, $g_{ij}$ and
$H_{ijk}$ are
unaffected to order $\beta$. Also note that we can interpret
the monopoles
as either strings in the space $(01234)$ or point
objects in the three-dimensional subspace $(0123)$.

The kinetic Lagrangian is obtained by replacing the expressions
for the
fields in \orderbeta\ into the string $\sigma$-model action
\stringact\ \footnote{$^*$}{Strictly speaking one must add to
\stringact\ a surface term to cancel the double derivative
terms in the action
\refs{\gibhp\gibhpone{--}\brih,\khumant,\khumonscat}\ however
the addition of
this term introduces only flat kinetic terms and thus presents
 no nontrivial
contribution to the metric on moduli space.}. Since \orderbeta\
is a solution
to
order $\beta$, the leading order terms in the action (after the
quasi-static part) are of order $\beta^2$. The
 $O(\beta)$ terms in the solution give $O(\beta^2)$ terms when
replaced in the
kinetic action. Collecting all $O(\beta^2)$ terms in
\stringact\ we
get the following kinetic Lagrangian density for the volume
term:
\eqn\kinlag{\eqalign{{\cal L}_{kin}=-{1\over 2\kappa^2}\Biggl(
&4\dot \phi\vec M\cdot\vec \nabla\phi
-e^{-2\phi}\partial_iM_j\partial_iM_j
-e^{-2\phi}M_k\partial_j\phi\left(\partial_jM_k-\partial_kM_j
\right)\cr
&+4M^2e^{-2\phi}(\vec \nabla\phi)^2
+2\partial_t^2e^{2\phi}-4\partial_t(\vec M\cdot\vec \nabla\phi)
-4\vec
\nabla\cdot(\dot\phi\vec M)\Biggr),\cr}}
where $\vec M\equiv -\sum_{n=1}^N{m_n\vec \beta_n\over |\vec x
 - \vec a_n(t)|}$.
Henceforth let $\vec X_n\equiv \vec x - \vec a_n(t)$.
The last three terms in \kinlag\ are time-surface or
space-surface terms
which vanish when integrated over the uncompactified four-space
$(0123)$.
The kinetic Lagrangian
$L_{kin}=\int d^3x{\cal L}_{kin}$ for monopole scattering
converges
everywhere. This can be seen simply by studying the limiting
behaviour
of $L_{kin}$ near each monopole. For a single monopole at $r=0$
with magnetic
charge $m$ and velocity $\beta$, we collect the logarithmically
divergent pieces
and find that they cancel:
\eqn\logdiv{{m\beta^2\over 2}\int r^2 drd\theta \sin\theta
d\phi
\left(-{1\over r^3} + {3\cos^2\theta\over r^3}\right)=0.}

We now specialize to the case of two identical monopoles of
magnetic
charge $m_1=m_2=m$ and velocities $\vec\beta_1$ and
$\vec\beta_2$.
Let the monopoles be located at $\vec a_1$ and $\vec a_2$.
Our moduli space consists of the configuration space of the
relative
separation vector $\vec a\equiv \vec a_2 - \vec a_1$.
The most general kinetic Lagrangian can be written as
\eqn\genkinlag{\eqalign{L_{kin}=&h(a)(\b1\cdot\b1+\b2\cdot\b2)
+p(a)\left(
(\b1\cdot\hat a)^2 + (\b2\cdot\hat a)^2\right)\cr
&+2f(a)\b1\cdot\b2 + 2g(a)(\b1\cdot\hat a)(\b2\cdot\hat a).\cr}}
Now suppose $\b1 = \b2 =\vec\beta$, so that \genkinlag\ reduces
to
\eqn\boostlag{L_{kin}=(2h+2f)\beta^2+(2p+2g)(\vec\beta\cdot\hat
a)^2.}
This configuration, however, represents the boosted solution of
 the
two-monopole static solution. The kinetic energy should
therefore be
simply
\eqn\cmke{L_{kin}={M_T\over 2}\beta^2,}
where $M_T=M_1+M_2=2M$ is the total mass of
the two-monopole solution. It then follows that the anisotropic
part of
\boostlag\ vanishes and we have
\eqn\hfpg{\eqalign{g+p&=0,\cr 2(h+f)&={M_T\over 2}.\cr}}
It is therefore sufficient to compute $h$ and $p$. This can be
done by
setting $\vec\beta_1=\vec\beta$ and $\vec\beta_2=0$.
The kinetic Lagrangian then reduces to
\eqn\rdkinlag{L_{kin}=h(a)\beta^2 +
p(a)(\vec\beta\cdot\hat a)^2.}
Suppose for simplicity
also that $\vec a_1=0$ and $\vec a_2=\vec a$ at $t=0$.
The Lagrangian density of the volume term in this case is given
by
\eqn\voltm{\eqalign{{\cal L}_{kin}&={-1\over 2\kappa^2}\Biggl(
{3m^3 e^{-4\phi}\over 2r^4}(\vec\beta\cdot\vec x)\left[
{\vec\beta\cdot\vec x\over r^3} + {\vec\beta\cdot(\vec x-\vec a)
\over |\vec x-\vec a|^3}\right] -
{e^{-2\phi}m^2\beta^2\over r^4}\cr
&-{e^{-4\phi}m^3\beta^2\over 2r^4}\left( {1\over r} +
{\vec x\cdot(\vec x-\vec a)\over |\vec x-\vec a|^3}\right) +
{e^{-6\phi}m^4\beta^2\over r^2}\left( {1\over r^4} + {1\over |
\vec x-\vec a|^4}
+ {2\vec x\cdot(\vec x-\vec a)\over r^3|\vec x-\vec a|^3}\right)
\Biggr).\cr}}

The integration of the kinetic Lagrangian density in \voltm\
over three-space
yields the kinetic Lagrangian from which the metric on moduli
space can be
read off. For large $a$, the nontrivial leading order
behaviour of the
components of the metric, and hence for the functions $h(a)$
 and $p(a)$, is
generically of order $1/a$. In fact, for Manton scattering of
BPS monopoles,
the leading order scattering angle is $2/b$ \mantwo, where $b$
is the impact
parameter. Here we restrict our computation to the leading
order
metric in moduli space. A tedious but straightforward
collection of $1/a$
terms in the Lagrangian yields
\eqn\leadi{{-1\over 2\kappa^2}{1\over a}
\int d^3x\left[ -{3m^4e^{-6\phi_1}
\over r^7}(\vec\beta\cdot\vec x)^2 + {m^3e^{-4\phi_1}\over r^4}
\beta^2+
{m^4e^{-6\phi_1}\over r^5}\beta^2 - {3m^5e^{-8\phi_1}\over r^6}
\beta^2
\right],}
where $e^{2\phi_1}\equiv 1+m/r$.
The first and third terms clearly cancel after integration over
three-space.
The second and fourth terms are spherically symmetric. A simple
integration
yields
\eqn\leadii{\int_0^\infty r^2dr \left( {e^{-4\phi_1}\over r^4}
-
{3m^2e^{-8\phi_1}\over r^6}\right)
=\int_0^\infty {dr\over (r+m)^2} - 3m^2\int_0^\infty {dr\over
(r+m)^4}=0.}
The $1/a$ terms therefore cancel, and the leading order metric
on moduli
space is flat. This implies that to leading order the dynamical
force is zero and the scattering is trivial, in agreement with
the test-soliton result.
In other words, there is no deviation from the initial
trajectories to
leading order in the impact parameter. Analogous computations
for elementary
strings in
$D=4$ \khugeo\ and fivebranes in $D=10$ \refs{\khumant,\fels}\
lead to the
same result of a flat metric. From $S \leftrightarrow T$
duality (see section
4) it follows
that the metric on moduli space for solitonic strings in $D=4$
is also flat.

We now address the scattering problem from the string
theoretic point of view. In particular, we calculate the string
four-point amplitude for the scattering of macroscopic winding
state
strings in the infinite winding radius limit. In this scenario,
we can
best approximate the soliton scattering problem considered
above but in the case of elementary strings in $D=4$.
We find that the Veneziano amplitude obtained also indicates
trivial
scattering in the large winding radius limit, thus providing
evidence
for the identification of the elementary strings with
infinitely
long macroscopic fundamental strings. The fivebrane analog of
this computation
awaits the construction of a fundamental fivebrane theory.
However, a vertex
operator representation of fivebrane solitons (and also of
string monopoles)
should in principle be possible.
The computation of the fivebrane Veneziano amplitude would then
represent a
dynamical test for string/fivebrane duality.

The scattering problem is set up in four dimensions, as the
kinematics
correspond essentially to a four dimensional scattering
problem, and
strings in higher dimensions generically miss each other anyway
\polc.
The precise compactification scheme is irrelevant to our
purposes.

The winding state strings then reside in four spacetime
dimensions
$(0123)$, with one of the dimensions, say $x_3$, taken to be
periodic with
period $R$, called the winding radius. The winding number $n$
describes
the number of times the string wraps around the winding
dimension:
\eqn\wind{x_3\equiv x_3 + 2\pi Rn,}
and the length of the string is given by $L=nR$. The integer
$m$, called
the momentum number of the winding configuration, labels the
allowed
momentum eigenvalues. The momentum in the winding direction is
thus
given by
\eqn\pthree{p^3={m\over R}.}
The number $m$ is restricted to be an integer so that the
quantum wave function
$e^{ip\cdot x}$ is single valued.
The total momentum of  each string can be written as the sum of
a
right momentum and a left momentum \eqn\tmoment{p^\mu=
p^\mu_R+p^\mu_L,}
where $p^\mu_{R,L}=(E,E\vec v,{m\over 2R}\pm nR)$,
$\vec v$ is the transverse velocity and $R$ is the winding
radius.
The mode expansion of the general
configuration $X(\sigma,\tau)$ in the winding direction
satisfying the two-dimensional wave equation
and the closed string boundary conditions can be written as the
sum of
right moving pieces and left moving pieces, each with the mode
expansion
of an open string \gresw~
\eqn\movers{\eqalign{X(\sigma,\tau)&=X_R(\tau -\sigma) +
X_L(\tau +\sigma)\cr
X_R(\tau -\sigma)&=x_R + p_R(\tau -\sigma) + {i\over 2}
\sum_{n=0} {1\over n}\alpha_n e^{-2in(\tau - \sigma)}\cr
X_L(\tau +\sigma)&=x_L + p_L(\tau +\sigma) + {i\over 2}
\sum_{n=0} {1\over n}\tilde\alpha_n e^{-2in(\tau + \sigma)}.\cr}}
The right moving and left moving components are then
essentially
independent parts with corresponding vertex operators, number
operators
and Virasoro conditions.

The winding configuration described by $X(\sigma,\tau)$
describes a
soliton string state. It is therefore a natural choice for us
to compare
the dynamics of these states with the soliton-like solutions of
the previous
sections (including the elementary solutions) in order to
determine whether we can identify the elementary string
solutions of
the supergravity field equations with infinitely long
fundamental
strings. Accordingly, we study the scattering of the winding
states in
the limit of large winding radius.

Our kinematic setup is as follows. We consider the scattering
of
two straight macroscopic strings in the CM frame with
winding number $n$ and momentum number $\pm m$
\refs{\gresw,\polc}.
The incoming momenta in the CM frame are given by
\eqn\imoment{\eqalign{p^\mu_{1R,L}&=(E,E\vec v,{m\over 2R}\pm
nR)\cr
p^\mu_{2R,L}&=(E,-E\vec v,-{m\over 2R}\pm nR).\cr}}
Let $\pm m'$ be the outgoing momentum number.
For the case of $m=m'$, the outgoing momenta are given by
\eqn\omoment{\eqalign{-p^\mu_{3R,L}&=(E,E\vec w,{m\over 2R}\pm
nR)\cr
-p^\mu_{4R,L}&=(E,-E\vec w,-{m\over 2R}\pm nR),\cr}}
where conservation of momentum and winding number have been
used and
where $\pm\vec v$ and $\pm\vec w$ are the incoming and outgoing
velocities of
the strings in the transverse $x-y$ plane. The outgoing momenta
winding numbers are not {\it a priori} equal to the initial
winding
numbers, but must add up to $2n$. Conservation of energy for
sufficiently large $R$ then results in the above answer. This
is also in
keeping with the soliton scattering nature of the problem (i.e.
the
solitons do not change ``shape" during a collision).

For now we have assumed no longitudinal excitation ($m=m'$).
We will later relax this condition to allow for such
excitation, but
show that our answer for the scattering is unaffected by this
possibility. It follows  from this condition that
$v^2=w^2$. For simplicity we take $\vec v=v\hat x$ and
$\vec w=v(\cos\theta\hat x+\sin\theta\hat y)$, and thus reduce
the
problem to a two-dimensional scattering problem.

As usual, the Virasoro conditions $L_0=\widetilde{L}_0=1$ must
hold, where
\eqn\vops{\eqalign{L_0&=N+\half (p^\mu_R)^2\cr
\widetilde{L}_0&=\widetilde
{N}+\half (p^\mu_L)^2 \cr}}
are the Virasoro operators \gresw\ and where $N$ and
$\widetilde{N}$ are the
number operators for the right- and left-moving modes
respectively:
\eqn\numbs{\eqalign{N&=\sum \alpha^\mu_{-n}\alpha_{n\mu}\cr
\widetilde{N}&=\sum \tilde\alpha^\mu_{-n}\tilde\alpha_{n\mu},
\cr}}
where we sum over all dimensions, including the compactified
ones.
It follows from the Virasoro conditions that
\eqn\evr{\eqalign{\widetilde{N}-N&=mn\cr
	E^2(1-v^2)&=2N-2+{({m\over 2R}+nR)}^2.\cr}}

In the following we set $n=1$ and consider for simplicity the
scattering
of tachyonic winding states. For our purposes, the nature of
the string
winding states considered is irrelevant. A similar calculation
for
massless bosonic strings or heterotic strings, for example,
will be
slightly more complicated, but will nevertheless exhibit the
same essential
behaviour. For tachyonic winding states we have
$N=\widetilde{N}=m=0$.
Equation \evr\ reduces to
\eqn\tevr{E^2(1-v^2)=R^2-2.}
The Mandelstam variables $(s,t,u)$ are identical for right and
left
movers and are given by
\eqn\mandlestam{\eqalign{s&=4\left[\x-2\right]\cr
t&=-2\left[\x\right](1+\cos\theta)\cr
u&=-2\left[\x\right](1-\cos\theta).\cr}}
It is easy to see that
$p_{iR}\cdot p_{jR}=p_{iL}\cdot p_{jL}$ holds
for this configuration so that the tree level 4-point function
reduces to the usual Veneziano amplitude for closed tachyonic
strings \polc
\eqn\veneziano{\eqalign{A_4&={\kappa^2\over 4}
B(-1-s/2,-1-t/2,-1-u/2)\cr
&=({\kappa^2\over 4}) {\Gamma(-1-s/2)\Gamma(-1-t/2)
\Gamma(-1-u/2)\over
\Gamma(2+s/2)\Gamma(2+t/2)\Gamma(2+u/2)}.\cr}}
This can be seen as follows. In the standard computation of the
four
point function for closed string tachyons, we rely on the
independence
of the right and left moving open strings. For the tachyonic
winding
state, we also separate the right and left movers with vertex
operators
given by $V_R=e^{ip_R\cdot x_R}$ and $V_L=e^{ip_L\cdot x_L}$
respectively
to arrive at the following expression for the amplitude
\eqn\afour{A_4={\kappa^2\over 4}\int d\mu_4(z)\prod_{i<j}
|z_i-z_j|^{p_{iR}\cdot p_{jR}} |z_i-z_j|^{p_{iL}\cdot p_{jL}}.}
{}From $p_{iR}\cdot p_{jR}=p_{iL}\cdot p_{jL}$, \afour\ reduces
to the
expression for the four-point amplitude of a nonwinding closed
tachyonic
string, from which the standard Veneziano amplitude in
\veneziano\ results.

To compare the implications of $A_4$ with the results of
the Manton calculation, we take $R\to\infty$. It is
convenient to define $x\equiv\x=s/4+2$, since the Mandelstam
variables can
be expressed solely in terms of $x$ and $\theta$. We now have
$A_4=A_4(x,\theta)$, which can be explicitly written as
\eqn\ampone{A_4=({\kappa^2\over 4})
{\Gamma(3-2x)\Gamma(-1+x(1+\cos\theta))\Gamma(-1+x(1-\cos\theta))\over
\Gamma(-2+2x)\Gamma(2-x(1+\cos\theta))\Gamma(2-x(1-\cos\theta))}.}
The problem reduces to studying $A_4$ in the limit $x\to\infty$.
We now use the identity $\Gamma(1-a)\Gamma(a)\sin\pi a=\pi$ to
rewrite
$A_4$ as
\eqn\amptwo{\eqalign{A_4=({\kappa^2\over 4\pi})&
\left[{\Gamma(-1+x(1+\cos\theta))
\Gamma(-1+x(1-\cos\theta))\over \Gamma(-2+2x)}\right]^2\cr
&\times\left({\sin(\pi x(1+\cos\theta))\sin(\pi x(1-\cos\theta))
\over\sin
2\pi x}\right).\cr}}
{}From the Stirling approximation $\Gamma(u)\sim\sqrt{2\pi}
u^{u-1/2}e^{-u}$
for large $u$, we obtain in the limit $x\to\infty$
\eqn\ampthree{\eqalign{A_4\sim&\left[{\left(x(1+\cos\theta)
\right)
^{x(1+\cos\theta)}
\left(x(1-\cos\theta)\right)^{x(1-\cos\theta)}\over (2x)^{2x}}
\right]^2\cr
&\times\left({\sin(\pi x(1+\cos\theta))
\sin(\pi x(1-\cos\theta))\over\sin
2\pi x}\right).\cr}}
Note that the exponential terms cancel automatically. From
\ampthree\ we
notice that the powers of $x$ in the first factor also cancel.
 $A_4$
then reduces in the limit $x\to\infty$ to
\eqn\amp{\eqalign{A_4\sim\left({1+\cos\theta\over 2}
\right)^{2x(1+\cos\theta)}
&\left({1-\cos\theta\over 2}\right)^{2x(1-\cos\theta)}\cr
&\times\left({\sin(\pi x(1+\cos\theta))
\sin(\pi x(1-\cos\theta))\over\sin
2\pi x}\right).\cr}}
The poles in the third factor in \amp\ are just the usual
 $s$-channel poles.
It follows from \amp\ that for $\theta\neq 0,\pi$~
$A_4 \to e^{-f(\theta)x}$ as $x\to\infty$,
where $f$ is some positive definite function of $\theta$.
Hence the 4-point function vanishes exponentially with the
winding radius
away from the poles.

In general, for finite $R$ and fixed $v$ the strings may
scatter into
longitudinally excited final states, {\it i.e.} states not
satisfying
the above assumption that $m'=m$. The $4$-point amplitude for
each
transition still vanishes exponentially with $R$.  A simple
counting
argument shows that the total number of possible final states
for a
given $R$ is bounded by a polynomial function of $R$. This
counting
argument proceeds as follows.

Without loss of generality, we may assume that our incoming
states have
$N=\widetilde{N}=m=0$ with fixed $R$ and $v$. We relax the
assumption of
no logitudinal excitation to obtain outgoing states with
nonzero $m$.
We still consider $n=1$ winding states for simplicity. Our
scattering
configuration can now be described by the incoming momenta
\eqn\imom{\eqalign{p^\mu_{1R,L}&=(E,E\vec v,\pm R)\cr
p^\mu_{2R,L}&=(E,-E\vec v,\pm R).\cr}}
and the outgoing momenta
\eqn\omom{\eqalign{-p^\mu_{3R,L}&=(E_1,E_1\vec w_1,
{m\over 2R}\pm R)\cr
-p^\mu_{4R,L}&=(E_2,-E_2\vec w_2,-{m\over 2R}\pm R).\cr}}
Note that in general $E_1$ and $E_2$ are not equal to $E$.
Without loss
of generality, we take $m$ to be positive. From
conservation of momentum, however, we have
\eqn\conserv{\eqalign{E_1+E_2&=2E\cr E_1\vec w_1&=E_2\vec w_2.\cr}}
It follows from the energy momentum relations for the ingoing and
outgoing momenta that
\eqn\enmom{\eqalign{E^2(1-v^2)&=R^2-2\cr
E_1^2(1-w_1^2)&=2N_1-2+\left({m\over 2R}+R\right)^2\cr
E_2^2(1-w_2^2)&=2N_2-2+\left(-{m\over 2R}+R\right)^2,\cr}}
where $N_1$ and $N_2$ are the number operators for the the right movers
of the outgoing states.

Subtracting the third equation in \enmom\ from the second equation and
using \conserv\ we obtain the relation
\eqn\nme{N_1-N_2+m=(E_1-E_2)E.}
{}From the first equation in \enmom\ it follows that $E$ is
bounded by
some multiple of $R$ for fixed $v$. It then follows from the
first
equation in \conserv\ that both $E_1$ and $E_2$ are bounded by
a
multiple of $R$. So from \nme\ we see that $N_1-N_2+m$ is
bounded by
some quadratic polynomial in $R$. We now add the last two
equations in
\enmom\ to obtain
\eqn\eenn{E_1^2(1-w_1^2)+E_2^2(1-w_2^2)=2N_1+2N_2+2R^2+
{m^2\over
2R^2}-4.}
The left hand side of \eenn\ is clearly bounded by a quadratic
polynomial in $R$. It follows that $N_1+N_2$ is also bounded by
a
quadratic polynomial, and that so are $N_1$ and $N_2$ and also,
then,
$N_1-N_2$. From the boundedness of $N_1-N_2+m$ it therefore
follows that
$m$ is bounded by a polynomial in $R$. Therefore
the total number of possible distinct excited states (numbered
by $m$)
is bounded by a polynomial in $R$. The above argument also goes
through
for the case of a nonzero initial momentum number. For each
transition,
however, one can show that the Veneziano amplitude is dominated
by an
exponentially vanishing function of $R$, from a calculation
entirely
analogous to the zero-longitudinal excitation case worked out
above.
Hence the total square amplitude of the scattering (obtained by
summing the square amplitudes of all possible transitions) is
still
dominated by a factor which vanishes exponentially with the
radius,
except at the poles at $\theta=0,\pi$ corresponding to forward
and
backward scattering, which are physically equivalent for
identical bosonic
strings. This is in agreement with the trivial
scattering found above and provides further evidence for the
identification of the elementary string with the fundamental
string.

The above argument can be repeated for any other type of
string, including
the heterotic string \grohmr. The kinematics differ slightly
from the
tachyonic case but the $4$-point function is still dominated by
an
exponentially vanishing factor in the large radius limit. Hence
the
scattering is trivial, again in agreement with the result found
above.

The Veneziano amplitude result in fact holds for arbitrary
incoming winding
states. A considerably more tedious calculation for the general
case
shows that in the large winding radius limit the outgoing
strings always
scatter trivially and with no change in their individual
winding numbers \khuwind.
In this limit, then, these states scatter as true solitons. It
would be interesting to see if this result holds for the full
quantum string
loop expansion.

\newsec{String Solitons and Supersymmetry}

In this section I discuss some recent results found in
\duffkr.
Most of the solitonic solutions found
so far break half of the spacetime supersymmetries
of the theory in which they arise. Examples
of string-like solitons
in this class are
the fundamental ($S$) string solution of \dabghr\ and the dual
($T$)
string solution of \dufkexst\ discussed in section 4 in the
context of string/string duality.

Interesting examples of solutions which break more than
half of the spacetime supersymmetries are
the double-instanton string soliton of \khubifb\ and the
octonionic string soliton of \hars. In this section I
 present new classes of string-like solutions which
arise in heterotic string theory toroidally compactified
to four dimensions. Connections are made between the
solution-generating subgroup of the
$T$-duality group and the number of spacetime
supersymmetries broken in the $N=4$ theory. Analogous
solutions are also seen to arise in $N=2$ and $N=1$
compactifications.

I adopt the following conventions for
$N=1$, $D=10$ heterotic string theory compactified to
$N=4$, $D=4$
heterotic string theory: $(0123)$ is the
four-dimensional spacetime, $z=x_2+ix_3=re^{i\theta}$,
$(456789)$ are
the compactified directions, $S=e^{-2\Phi} + ia=S_1+iS_2$, where
$\Phi$ and $a$ are the four-dimensional dilaton and axion and
\eqn\tdefn{\eqalign{T^{(1)}&=T^{(1)}_1+iT^{(1)}_2=\sqrt{{\rm det}
g_{mn}}-iB_{45}, \quad\quad m,n=4,5,\cr
T^{(2)}&=T^{(2)}_1+iT^{(2)}_2=\sqrt{{\rm det} g_{pq}}-iB_{67},
\quad\quad p,q=6,7,\cr
T^{(3)}&=T^{(3)}_1+iT^{(3)}_2=\sqrt{{\rm det} g_{rs}}-iB_{89},
\quad\quad r,s=8,9\cr}}
are the moduli. Throughout this section,
and unless specified otherwise,
 I assume dependence only on the
coordinates $x_2$ and $x_3$ (i.e. $x^1$ remains a
Killing direction), and that no other moduli than the ones above
are nontrivial.

The canonical four-dimensional bosonic action for the
above compactification ansatz in the
gravitational sector can be written in terms of $g_{\mu\nu}$
($\mu,\nu=0,1,2,3$), $S$ and $T^{(a)}, a=1,2,3$ as
\eqn\sfour{\eqalign{S_4=\int d^4x \sqrt{-g}\biggl(&R-
{g^{\mu\nu}\over
2S^2_1}\partial_\mu S \partial_\nu \bar S \cr &-
{g^{\mu\nu}\over
2T^{(1)^2}_1}\partial_\mu T^{(1)}
\partial_\nu \bar T^{(1)} - {g^{\mu\nu}\over
2T^{(2)^2}_1}\partial_\mu T^{(2)}
\partial_\nu \bar T^{(2)} - {g^{\mu\nu}\over
2T^{(3)^2}_1}\partial_\mu T^{(3)}
\partial_\nu \bar T^{(3)} \biggr).\cr}}
A solution for this action for $S=1$ ($\Phi=a=0$) is given
by
the metric
\eqn\metsol{ds^2=-dt^2+dx_1^2 + T^{(1)}_1 T^{(2)}_1 T^{(3)}_1
(dx_2^2+dx_3^2),}
where three cases with different nontrivial $T$ duality arise
 depending on
the
number $n$ of nontrivial $T$ moduli:
\eqn\tsol{\eqalign{n&=1: \qquad\qquad T^{(1)}=-{1\over 2\pi}\ln
{z\over r_0},
\qquad T^{(2)}=T^{(3)}=1,\cr
n&=2: \qquad\qquad T^{(1)}=T^{(2)}=-{1\over 2\pi}\ln
{z\over r_0},
\qquad T^{(3)}=1,\cr
n&=3: \qquad\qquad T^{(1)}=T^{(2)}=T^{(3)}=-{1\over 2\pi}
\ln {z\over
r_0}.\cr}}
In each of the expressions for $T^{(a)}$, $z$ may be replaced by
$\bar
z$ independently (i.e. there is a freedom in changing
 the sign of the
axionic part of the modulus).
Note that the $n=1$ case is simply the dual
string solution of \dufkexst.
Since $S_4$ has manifest $SL(2,R)$ duality in each of the moduli
(broken to $SL(2,Z)$ in string theory), we
can generate from the $n=2$ case an $SL(2,Z)^2$ family of
 solutions
and from the $n=3$ case an $SL(2,Z)^3$ family of solutions.
Note that the full $T$ duality group in all three cases
remains $O(6,22;Z)$, but that the subgroup with nontrivial
action on the particular solutions (or the solution-generating
subgroup referred to above) for $n=1,2,3$ is given
by $SL(2,Z)^n$ (see \refs{\fkp,\fp}).

{}From the ten-dimensional viewpoint, the $n=3$ solution, for
example, can be rewritten in the string sigma-model metric
 frame as
\eqn\tend{\eqalign{e^{2\phi}&=(-{1\over 2\pi}
\ln {r\over r_0})^3,\cr
ds^2&=-dt^2+dx_1^2 + e^{2\phi} (dx_2^2+dx_3^2) +
e^{2\phi/3} (dx_4^2 + ...+ dx_9^2),\cr
B_{45}&=\pm B_{67}=\pm B_{89}=\pm {\theta\over 2\pi},\cr}}
where $\phi$ is the ten-dimensional
dilaton.

The solution \tsol\ can in fact be generalized to include an
arbirary number of string-like sources in each $T^{(i)}$
\eqn\threet{\eqalign{
ds^2&=-dt^2+dx_1^2 + T^{(1)}_1 T^{(2)}_1 T^{(3)}_1
(dx_2^2+dx_3^2) \cr
 T^{(1)}& =
-{1\over 2\pi} \sum_{j=1}^M m_j\ln {(z-b_j)\over r_{j0}} ,\cr
T^{(2)}& =
-{1\over 2\pi} \sum_{k=1}^P p_k\ln {(z-c_k)\over r_{k0}} ,\cr
T^{(3)}& =
-{1\over 2\pi} \sum_{l=1}^Q q_l\ln {(z-d_l)\over r_{l0}} ,\cr}}
where  $M, P$ and $Q$ are
arbitrary numbers of string-like solitons in
$T^{(1)}, T^{(2)}$ and $T^{(3)}$ respectively
each with arbitrary location $b_j, c_k$ and $d_l$ locations in the
complex $z$-plane and arbitrary
winding number $m_j, p_k$ and $q_l$
respectively. The solutions with
 $1,2$ and $3$ nontrivial $T$ fields break $1/2, 3/4$ and $7/8$
of the spacetime supersymmetries respectively. Again, one can
make the replacement $z\to \bar z$ independently
in each of the moduli, so that each $T^{(i)}$ is either
analytic or anti-analytic in $z$.

The above solutions for the massless fields in
the
gravitational sector when combined with a Yang-Mills field
 given by
$A_M^{PQ}=\Omega_M^{PQ}=\omega_M^{PQ} \pm 1/2 H_M{}^{PQ}$ (the
usual expedient of equating the
gauge to the generalized connection) solve the tree-level
supersymmetry equations of the heterotic string for zero fermi
fields
and can be argued to be exact solutions of heterotic string
theory \refs{\calhs,\dufkexst}.
The supersymmetry equations in $D=10$ are given by
\eqn\sseq{\eqalign{\delta\psi_M&=\left(\partial_M+{\textstyle
{1\over
4}}\Omega_{MAB}
\Gamma^{AB}\right)\epsilon=0,\cr
\delta\lambda&=\left(\Gamma^A\partial_A\phi-{\textstyle{1\over
12}}
H_{ABC}\Gamma^{ABC}\right)\epsilon=0,\cr
\delta\chi&=F_{AB}\Gamma^{AB}\epsilon=0, \cr}}
where $A,B,C,M=0,1,2,...,9$ and
where $\psi_M,\ \lambda$ and $\chi$ are the gravitino, dilatino
and
gaugino
fields. The Bianchi identity is given by
\eqn\bianchi{dH={\alpha'\over 4} \left({\rm tr} R\wedge R-
{1\over
30}{\rm Tr}
F\wedge F\right),}
and is satisfied automatically for this ansatz. We further claim
that
the $n=1,2,3$ solutions break $1/2, 3/4$ and $7/8$ of the
spacetime
supersymmetries respectively. We will show this to be true for
the most general case of $n=3$.

$\delta\lambda=0$ follows from scaling, since the dilaton can
be
written as the sum of three parts (the moduli) each of which
produces
a contribution which cancels against the contribution of the
$H$ term
coming from the appropriate
four-dimensional subspace. In other words, each of the
subspaces
$(2345)$, $(2367)$ and $(2389)$ effectively contains a
four-dimensional axionic  (anti) instanton
\refs{\reyone,\khuinst,\khumonin}\ with the appropriate (anti)
self-duality in the
generalized connection in the respective
subspace, depending on whether the corresponding
modulus is analytic or anti-analytic in $z$.
 Another way of saying this is that there are three
independent parts of $\delta\lambda$, each of which vanishes
as in the simple $n=1$
case, for the appropriate chirality choice of $\epsilon$ in the
respective four-dimensional subspace.

$\delta\psi_M=0$ is a little more subtle. For the $n=1$
case,
the generalized connection is an
instanton \refs{\khuinst,\khumonin}, and for constant
chiral spinor $\epsilon$ with
chirality
in the
four-space of the instanton opposite to that of the instanton
(e.g.
negative for instanton and positive for anti-instanton), it is
easy
to show that
$\Omega_M^{AB}\Gamma_{AB}\epsilon=0$. In the more general
$n=3$ case, we
proceed as follows. It is sufficient to show that
$\delta\psi_M=0$
for $M=2$ and $M=4$
(i.e. for a spacetime and for a compactified index),
as for $M=0,1$ the supersymmetry variation is trivial,
while for the rest of the indices the arguments are identical
to one of the above two representative cases. For $M=2$
this
can be written out explicitly as
\eqn\psitwo{\eqalign{4\delta\psi_2=&\left({1\over 3}
\omega_2^{23}
\Gamma_{23} +
\omega_2^{24} \Gamma_{24} + \omega_2^{25} \Gamma_{25} -
{1\over 2}
H_2{}^{45}
\Gamma_{45}\right) \epsilon \cr
+  &\left({1\over 3} \omega_2^{23} \Gamma_{23} +
\omega_2^{26} \Gamma_{26} + \omega_2^{27} \Gamma_{27} -
{1\over 2}
H_2{}^{67}
\Gamma_{67}\right) \epsilon \cr
+  &\left({1\over 3} \omega_2^{23} \Gamma_{23} +
\omega_2^{28} \Gamma_{28} + \omega_2^{29} \Gamma_{29} -
{1\over 2}
H_2{}^{89}
\Gamma_{89}\right) \epsilon .\cr}}
Each line in \psitwo\ acts on only a four-dimensional component
of $\epsilon$ and can be shown to exactly correspond to the
contribution of the supersymmetry equation of a single $n=1$
axionic instanton.
So in effect the configuration carries three such instantons
in the
generalized curvature in the spaces $(2345)$, $(2367)$ and
$(2389)$.
Therefore for the appropriate chirality of the four-dimensional
components of $\epsilon$ (depending on the choices of analyticity
of the $T$ fields),
$\delta\psi_2=0$. Since we are making three such choices, $1/8$
of the spacetime supersymmetries are preserved and $7/8$
are broken. Another, perhaps
simpler, way to understand this is to write
$\epsilon=\epsilon_{(01)}\otimes \epsilon_{(23)}\otimes
\epsilon_{(45)}\otimes
\epsilon_{(67)}\otimes \epsilon_{(89)}$. Then the chiralities of
$\epsilon_{(45)}, \epsilon_{(67)}$ and $\epsilon_{(89)}$ are all
correlated with that
of $\epsilon_{(23)}$, so it follows that $7/8$ of the
supersymmetries
are broken.

We also need to check $\delta\psi_4=0$. In this case, it is
easy to
show that the whole term reduces exactly to the contribution of
a single $n=1$ axionic instanton:
\eqn\psifour{4\delta\psi_4= \left(\omega_4^{42} \Gamma_{42} +
\omega_4^{43} \Gamma_{43} - {1\over 2} H_4{}^{25}
\Gamma_{25} - {1\over 2} H_4{}^{35}
\Gamma_{35}\right)\epsilon =0,}
as in this case there is only the contribution of the instanton
in the $(2345)$ subspace. $\delta\psi_4$ then vanishes for the
same chirality choice of $\epsilon$ as in the paragraph
above.

There remains to show that $\delta\chi=0$. This can be easily
seen by
noting that, as in the $\delta\psi_M$ case, the term
$F_{23}\Gamma^{23}$ splits into three equal pieces, each of
which
combines with the rest of a $D=4$ instanton
(since the Yang-Mills connection is equated to the generalized
connection and is also effectively an instanton in each of the
three
four-dimensional subspaces) to give a zero contribution for
the same chirality choices in the four-dimensional
subspaces as above.

For the $n=2$ case, it is even easier to show that $3/4$ of the
supersymmetries are broken. Tree-level neutral versions ($A_M=0$) of these
solutions also follow immediately and reduce to \metsol\
and \tsol\ on compactification to $D=4$, where, of course,
the same degree of supersymmetry breaking for each
class of solutions may be verified
directly. Henceforth we will consider only neutral solutions.

It turns out that these solutions generalize even further to
solutions which include a nontrivial $S$ field. The net result
of
adding a nontrivial $S$ (with $SL(2,Z)$ symmetry) is to break
half again of the remaining spacetime supersymmetries
preserved by the corresponding $T$ configuration with trivial
$S$, except for the case of $n=3$ nontrivial moduli, which is a bit
more subtle and will be discussed below.
In particular, the simplest solution of the action
\sfour\ with one nontrivial $S$ and three nontrivial $T$
moduli has the form
\eqn\sthreet{\eqalign{ds^2&=-dt^2+dx_1^2
+ S_1  T_1^{(1)}  T_1^{(2)}  T_1^{(3)} (dx_2^2 + dx_3^2),\cr
S&=T^{(1)}=T^{(2)}=T^{(3)}=-{1\over 2\pi} \ln {z\over r_0},\cr}}
where again we have an $SL(2,Z)$ symmetry in $S$ and
in each of the
$T$ fields.

It is interesting to note that the real parts of the $S$ and
$T$ fields can
be arbitrary as long as they satisfy the box equation in
the two-dimensional subspace $(23)$. In particular, each can be
generalized to multi-string configurations independently,
with arbitrary number of strings each with
arbitrary winding number. The
corresponding imaginary part can most easily be found by going
to ten dimensions, where the corresponding $B$-field follows
from the modulus. So there is nothing special about the choice
$\ln z$. It is merely the simplest case.

The ten-dimensional form
of the most general solution can be written in the string
sigma-model metric frame as
\eqn\sthreetten{\eqalign{
ds^2&=e^{2\Phi}(-dt^2+dx_1^2) + e^{2(\sigma_1 + \sigma_2 +
\sigma_3)} (dx_2^2 + dx_3^2) \cr
 &+ e^{2\sigma_1} (dx_4^2 + dx_5^2)
+  e^{2\sigma_2} (dx_6^2 + dx_7^2) +
 e^{2\sigma_3} (dx_8^2 + dx_9^2), \cr
S&=e^{-2\Phi} + ia = -{1\over 2\pi} \sum_{i=1}^N n_i
\ln {(z-a_i)\over r_{i0}} ,\cr
T^{(1)}&=e^{2\sigma_1} - iB_{45} =
-{1\over 2\pi} \sum_{j=1}^M m_j\ln {(z-b_j)\over r_{j0}} ,\cr
T^{(2)}&=e^{2\sigma_2} - iB_{67} =
-{1\over 2\pi} \sum_{k=1}^P p_k\ln {(z-c_k)\over r_{k0}} ,\cr
T^{(3)}&=e^{2\sigma_3} - iB_{89} =
-{1\over 2\pi} \sum_{l=1}^Q q_l\ln {(z-d_l)\over r_{l0}} ,\cr
\phi&=\Phi +\sigma_1 + \sigma_2 + \sigma_3,\cr}}
where $\phi$ is the ten-dimensional dilaton, $\Phi$ is the
four-dimensional dilaton, $\sigma_i$ are the metric
moduli, $a$ is the axion in the
four-dimensional subspace $(0123)$ and $N, M, P$ and $Q$ are
arbitrary numbers of string-like solitons in
$S, T^{(1)}, T^{(2)}$ and $T^{(3)}$ respectively
each with arbitrary location $a_i, b_j, c_k$ and $d_l$ in the complex
$z$-plane and
arbitrary winding number $n_i, m_j, p_k$ and $q_l$
respectively. Again one can replace $z$ by $\bar z$ independently
in $S$ and in each of the moduli.

The solutions with nontrivial $S$ and $0, 1$ and $2$ nontrivial
$T$ fields preserve $1/2, 1/4$  and $1/8$ spacetime supersymmetries
respectively.
This follows from the fact that the nontrivial $S$ field breaks
half of the remaining supersymmetries by imposing a chirality
choice on the spinor $\epsilon$ in the
$(01)$ subspace of the ten-dimensional space.
The solution with nontrivial $S$
and $3$ nontrivial $T$ fields breaks $7/8$ of the spacetime
supersymmetries for one chirality choice of $S$, and all the
spacetime supersymmetries for the other. This can be seen as follows:
the three nontrivial $T$ fields, when combined with an overall
chirality choice of the Majorana-Weyl spinor in ten dimensions,
impose a chirality choice on $\epsilon_{01}$. If this choice
agrees with the chirality choice imposed by $S$, then no
more supersymmetries are broken, and so $1/8$ are preserved
(or $7/8$ are broken). When these two choices are not identical,
all the supersymmetries are broken, although the ansatz remains
a solution to the bosonic action.

A special case of the above generalized $S$ and $T$
 solutions is the one
with nontrivial $S$ and only one nontrival $T$.
This is in fact a ``dyonic'' solution which interpolates
between the fundamental $S$ string of \dabghr\ and the
dual $T$ string of \dufkexst. It turns out that in going to
higher
dimensions, one still has a solution even if the box
equation covers the whole transverse four-space $(2345)$
(the remaining four directions are flat even in $D=10$, as
$\sigma_2=\sigma_3=0$).
The $D=10$ form in fact reduces to a $D=6$ dyonic solution
($i=2,3,4,5$)
\eqn\dyon{\eqalign{\phi&=\Phi_E + \Phi_M, \cr
ds^2&=e^{2\Phi_E} (-dt^2+dx_1^2)
+ e^{2\Phi_M} dx_i dx^i,\cr
e^{-2\Phi_E}&=1+{Q_E\over y^2},\qquad\qquad
e^{2\Phi_M}=1+{Q_M\over y^2},\cr
H_3&=2Q_M\epsilon_3,\qquad\qquad
e^{-2\phi}{}{} ^* H_3=2Q_E\epsilon_3\cr}}
for the special case of a single electric and single
magnetic charge at $y=0$.
Again this solution generalizes to one with an arbitrary
number of arbitrary (up to dyonic
quantization conditions) charges
at arbitrary locations in the transverse four-space.
 For $Q_M=0$ \dyon\ reduces to the solution of
\dabghr\ in $D=6$, while for $Q_E=0$ \dyon\ reduces to the
$D=6$ dual string of \duflblacks\
(which can be obtained from the fivebrane soliton \duflfb\
simply
by compactifying four flat directions). This solution breaks
$3/4$ of the spacetime supersymmetries.
The self-dual limit $Q_E=Q_M$ of this solution has already been
found in \duflblacks\ in the context of $N=2$, $D=6$ self-dual
supergravity, where the solution was shown to
break $1/2$ the spacetime
 supersymmetries. This corresponds precisely to breaking
$3/4$ of the spacetime supersymmetries in the
non self-dual theory shown here \refs{\prep,\duffkr}.

Finally, one can generalize the dyonic solution to the
following solution in $D=10$:
\eqn\bibi{\eqalign{
ds^2 &=e^{2\Phi_E}(-dt^2+dx_1^2)
+e^{2\Phi_{M1}}\delta_{ij} dx^i dx^j
+e^{2\Phi_{M2}}\delta_{ab} dx^a dx^b, \cr
\phi &=\Phi_E + \Phi_{M1} + \Phi_{M2},\qquad\qquad
\Phi_E =\Phi_{E1} + \Phi_{E2}, \cr
e^{2\Phi_{E1}} \Box_1 \ e^{-2\Phi_{E1}} &=
e^{2\Phi_{E2}} \Box_2 \ e^{-2\Phi_{E2}}=
e^{-2\Phi_{M1}} \Box_1 \ e^{2\Phi_{M1}}=
e^{-2\Phi_{M2}} \Box_2 \ e^{2\Phi_{M2}}=0, \cr
B_{01} &=e^{2\Phi_E},\qquad
H_{ijk}=2\epsilon_{ijkm}\partial^m \Phi_{M1},\qquad
H_{abc}=2\epsilon_{abcd}\partial^d \Phi_{M1},\cr}}
where $i,j,k,l=2,3,4,5$, $a,b,c,d=6,7,8,9$, $\Box_1$ and
$\Box_2$
represent the Laplacians in the subspaces $(2345)$ and
$(6789)$ respectively and $\phi$ is the ten-dimensional
dilaton. This solution with all fields nontrivial
 breaks $7/8$ of the
spacetime supersymetries. For $\Phi_{E2}=\Phi_{M2}=0$ we
recover the dyonic solution \dyon\ which breaks $3/4$
of the supersymmetries, for $\Phi_{E1}=\Phi_{E2}=0$
we recover the double-instantaon solution of \khubifb\ which
also breaks
$3/4$ of the supersymmetries, while for $\Phi_{M1}=\Phi_{M2}=0$
we obtain the dual of the double-instanton solution, and which,
however, breaks only $1/2$ of the supersymmetries.

It turns out that most of the above solutions that break
$1/2, 3/4, 7/8$ or all of the spacetime supersymmetries in
$N=4$ have analogs in $N=1$ or $N=2$
that break only $1/2$ the spacetime supersymmetries.

For simplicity, let us consider the case of $N=1$,
as the $N=2$ case is similar.
It turns out that the number of nontrivial
$T$ fields with the same analyticity
and the inclusion of a nontrivial $S$ field with the same analyticity
does
not affect
the number of supersymmetries broken, as in the supersymmetry
equations the contribution of each field is independent. In
particular, the presence of an additional field produces no
new condition on the chiralities, so that the number of
supersymmetries broken is the same for any number of fields, provided
the fields have the same analyticity or anti-analyticity in $z$,
corresponding to different chirality choices on the four-dimensional
spinor. This can be seen as follows below.

The supersymmetry transformations in $N=1$ for nonzero
metric, $S$ and moduli fields are given by
\refs{\sg,\sgone}\
\eqn\ssnone{\eqalign{\delta\psi_{\mu L}&=
\left(\partial_\mu + {1\over 2}
\omega_{\mu mn}\sigma^{mn}\right) \epsilon_L
-{\epsilon_L\over 4}\left( {\partial G\over \partial z_i}
\partial_\mu z_i -{\partial  G\over \partial \bar z_i}
\partial_\mu \bar z_i \right)=0, \cr
\delta \chi_{iL}&={1\over 2} \hat D z_i \epsilon_R=0,\cr}}
where $\omega$ is the spin connection,
$\sigma^{mn}=(1/4)[\gamma^m,\gamma^n]$,
$\epsilon_{L,R}=(1/2)(1\pm \gamma^5)\epsilon$,
$\hat D=\gamma^\mu D_{mu}$, $z_i=S, T^{(1)}, T^{(2)}, T^{(3)}$,
and where
\eqn\gpot{G= -\ln (S+\bar S) - \sum_{j=1}^3 \ln (T^{(j)}+
\bar T^{(j)}).}
Consider the case of a single
nontrivial $T=T^{(1)}$ field (i.e. the dual string)
in $N=1, D=4$
\eqn\onet{\eqalign{ds^2&=-dt^2 + dx_1^2 + T_1 (dx_2^2
+ dx_3^2),\cr
T&=T_1 + iT_2=-{1\over 2\pi} \ln z.}}
Then it is easy to show that this configuration breaks
precisely half the spacetime supersymmetries of \ssnone\
by imposing two conditions on the components of $\epsilon$.
A quick check shows that the presence of additional nontrivial
$S$ and $T$ fields with the same analyticity behaviour
also lead to solutions of \ssnone, and this scenario
generalizes to multi-string solutions. The number, location and
winding numbers of the multi-string solitons is not relevant,
but the fact the fields have $\ln z$ or $\ln \bar z$ behaviour is.
Provided the $S$ and various $T$ fields all have the same
analyticity (i.e. either all analytic or all anti-analytic) in $z$,
then no new chirality choice is imposed by the addition of more fields.
This can
be seen simply from the fact that the spin connection and
potential $G$ both scale logarithmically with the fields,
while $\delta \chi_{iL}$ is satisfied in the identical
manner for each $i$. Unlike the
$N=4$ case, the presence of these additional fields
produces no new conditions on $\epsilon$, as the
supersymmetry variations act on $\epsilon$
in precisely the same manner for all the fields. It follows
then that the $N=1$ analogs of these particular $N=4$ solutions discussed
above break only half the spacetime supersymmetries in $N=1$,
and in some sense are realized naturally as stable
solitons only
in this context. When at least one of the fields, either $S$
or one of the $T$ fields, has a different analyticity behaviour
from the rest, opposite chirality conditions are imposed on
$\epsilon$, and no supersymmetries are preserved.

 A similar analysis can be done in the
$N=2$ case, at least for those solutions which can arise in
$N=2$.

The relationship between duality and supersymmetry is
also discussed in \refs{\baktwo,\bko}.

\newsec{Future Directions}

As mentioned in the introduction, an important possible
application of the instanton solutions in string theory
(in particular the axionic instanton discussed in this paper)
is the exploration of vacuum tunnelling in string theory. The
true stringy analogs of instanton computations in field theory
probably arise within the context of string field theory
\refs{\senzwione,\senzwitwo}. However, it is possible to
perform simpler computations in the low-energy
supergravity limit of string theory \reyt. Another possibility
is to try to obtain vertex-operator representations of states
corresponding to string instanton solutions and compute
transition amplitudes between vacuum states in the low-energy
limit. Comparisons with analogous results in point field theory
may then be quite illuminating.

The construction of vertex operators may also be useful as a
dynamical
test of the various dualities discussed in this paper. For
example, a vertex operator representation of
fivebranes would allow us to repeat the Veneziano amplitude
calculation of section 6 for fivebranes
and again compare with expectations
from the Manton metric on moduli space. As in the string
case, one would still expect a vanishing leading order dynamical
force in the limit of infinitely long fivebranes. A similar
result should also hold for the monopoles considered in this
paper.

The fact that these extremal $\alpha=\sqrt{3}$ black
holes/monopoles scatter trivially to leading order in the impact
parameter
is in direct contrast to not only field-theoretic analogs, such
as the scattering of BPS monopoles (as pointed out in section 6),
but also to analogous computations in general relativity
(see, e.g., \fere, where the metric on moduli space is computed
for extreme Reisnner-Nordstrom  black holes and is found to
be non-flat). Whether this feature is intrinsically stringy or
not is not entirely clear. However, the stark contrast with
results in generic Einstein-Maxwell-Higgs theories suggests a
possible test for string theory as the correct theory of quantum
gravity.

Finally, given the encouraging success of the application of
stringy techniques in statistical physics and QCD,
the question arises as to whether stringy methods may be
useful in relativity. An open problem in this regard
is the representation of the more astrophysically
realistic Schwarzschild black hole as
a conformal field theory. Such a representation would certainly
lead to important insights into the nature of string
theory as a theory of quantum gravity, but may in addition
lead to a greater understanding of the nature of singularities
in relativity. A (slightly) less ambitious project is to try to
adapt stringy scattering techniques, in conjunction with the
Manton approach and the various dualities
\refs{\hor,\kir,\bakone}, to analyze the low-velocity
interactions of Schwarzschild black holes.

\newsec{Acknowledgements}

I would like to thank Mike Duff, Sergio
Ferrara, Ruben Minasian and Joachim Rahmfeld for
collaboration and for helpful discussions.

\listrefs
\end